\documentclass[aps,prl,twocolumn,10pt,superscriptaddress]{revtex4-2}

\usepackage{mathrsfs}
\usepackage{bbm}
\usepackage{amsmath}
\usepackage{amssymb}
\usepackage{amsthm}
\usepackage{graphicx}
\usepackage{MnSymbol}
\usepackage{mathtools}
\usepackage[italicdiff]{physics}
\usepackage{dsfont}

\makeatletter
\usepackage{hyperref}
\usepackage{color}
\definecolor{supcol}{RGB}{10,50,180}
\definecolor{eqcol}{RGB}{220,10,100}
\hypersetup{
	colorlinks,
	citecolor=supcol,
	linkcolor=eqcol,
	urlcolor=supcol
}

\allowdisplaybreaks

\DeclareMathOperator{\mvar}{Var}

\newcommand{\mca}{\mathcal}

\newcommand{\msf}{\mathsf}

\newcommand{\mds}{\mathds}
\newcommand{\msc}{\mathscr}

\newcommand{\sectionprl}[1]{{\em #1}\/---}

\begin{document}
\title{Universal Precision Limits in General Open Quantum Systems}

\author{Tan Van Vu}
\email{tan.vu@yukawa.kyoto-u.ac.jp}
\affiliation{Center for Gravitational Physics and Quantum Information, Yukawa Institute for Theoretical Physics, Kyoto University, Kitashirakawa Oiwakecho, Sakyo-ku, Kyoto 606-8502, Japan}

\author{Ryotaro Honma}
\affiliation{Center for Gravitational Physics and Quantum Information, Yukawa Institute for Theoretical Physics, Kyoto University, Kitashirakawa Oiwakecho, Sakyo-ku, Kyoto 606-8502, Japan}

\author{Keiji Saito}
\affiliation{Department of Physics, Kyoto University, Kyoto 606-8502, Japan}

\date{\today}

\begin{abstract}
The intuition that the precision of observables is constrained by thermodynamic costs has recently been formalized through thermodynamic and kinetic uncertainty relations. While such trade-offs have been extensively studied in Markovian systems, corresponding constraints in the non-Markovian regime remain largely unexplored. In this Letter, we derive universal bounds on the precision of generic observables in open quantum systems that interact with their environments at arbitrary coupling strengths and are subjected to two-point measurements. By introducing an asymmetry term that quantifies the disparity between forward and backward processes, we show that the relative fluctuation of any time-antisymmetric current is constrained not only by entropy production but also by this asymmetry. For general observables, we further prove that their relative fluctuation is always bounded from below by a generalized activity term that characterizes environmental changes. These results establish a comprehensive framework for understanding the fundamental limits of precision in a broad class of open quantum systems, beyond the traditional Markovian setting.
\end{abstract}

\maketitle

\sectionprl{Introduction}Understanding the fundamental limits of precision is essential for the development of quantum machines such as sensors, clocks, and heat engines.
It is intuitive that high precision cannot be achieved for free and some cost must inevitably be paid.
In recent years, this intuition has been rigorously formalized through a class of uncertainty relations, initially developed for classical Markov jump processes \cite{Horowitz.2020.NP}.
The thermodynamic uncertainty relation (TUR) states that achieving high precision of currents---quantified as the squared mean divided by the variance---requires a corresponding increase in dissipation \cite{Barato.2015.PRL,Gingrich.2016.PRL,Horowitz.2017.PRE,Hasegawa.2019.PRE,Koyuk.2020.PRL,Vo.2022.JPA}.
Complementarily, the kinetic uncertainty relation (KUR) indicates that improving the precision of counting observables, defined in the same way, necessitates increased jump activity \cite{Garrahan.2017.PRE,Terlizzi.2019.JPA,Prech.2025.PRX}.
These relations not only reveal the costs associated with enhancing precision in small, fluctuating systems, but also carry significant implications for nonequilibrium physics \cite{Pietzonka.2018.PRL,Hartich.2021.PRL,Li.2019.NC,Manikandan.2020.PRL,Vu.2020.PRE}.

In the quantum regime, uncovering trade-off relations between precision and thermodynamic costs becomes significantly more nontrivial due to uniquely quantum features such as coherence and entanglement.
It has been shown that classical uncertainty relations---namely, the TUR and KUR---can be violated in quantum systems \cite{Agarwalla.2018.PRB,Ptaszynski.2018.PRB,Brandner.2018.PRL,Liu.2019.PRE,Saryal.2019.PRE,Cangemi.2020.PRB,Friedman.2020.PRB,Kalaee.2021.PRE,Menczel.2021.JPA,Bret.2021.PRE,Sacchi.2021.PRE,Lu.2022.PRB,Gerry.2022.PRB,Das.2023.PRE,Manzano.2023.PRR,Singh.2023.PRA,Farina.2024.PRE}.
For Markovian quantum dynamics, several extensions of these relations have been proposed, revealing how quantum coherence can enhance the precision of observables \cite{Guarnieri.2019.PRR,Carollo.2019.PRL,Hasegawa.2020.PRL,Miller.2021.PRL.TUR,Vu.2022.PRL.TUR,Prech.2025.PRL,Vu.2025.PRXQ,Macieszczak.2024.arxiv,Kwon.2025.CP,Moreira.2025.PRE,Yunoki.2025.arxiv,Nishiyama.2025.arxiv,Yoshimura.2026.PRL,Landi.2024.PRXQ}.
These studies demonstrate that quantum effects can relax classical bounds, enabling high precision even at low thermodynamic costs.
In the context of fermionic transport, precision bounds for particle currents beyond the Markovian regime have also been explored, highlighting the crucial role of coherent dynamics in enhancing precision \cite{Palmqvist.2025.PRL,Blasi.2025.arxiv,Brandner.2025.PRL}.
As a result, coherence and entanglement have been widely recognized as essential ingredients in modifying precision trade-offs.
Nevertheless, the ultimate limits that quantify how thermodynamic costs and quantum features {\it jointly} constrain the precision of trajectory observables in the strong-coupling, non-Markovian regime remain largely unexplored.
This gap motivates the development of precision bounds that apply to arbitrary system-environment couplings and extend beyond the limitations of Markovian approximations \cite{Hasegawa.2021.PRL,Ishida.2025.PRE,QTRoadmap.2026.QST}.

\begin{figure}[b]
\centering
\includegraphics[width=1\linewidth]{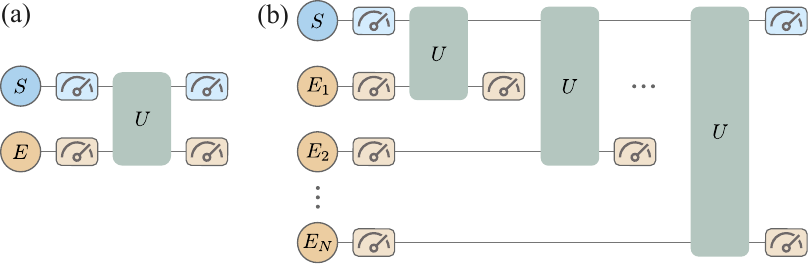}
\protect\caption{Schematic illustration of general open quantum systems interacting with uncorrelated environments and subjected to two-point measurements. (a) The system and environment evolve under a single unitary transformation, with measurements performed at the initial and final times. (b) The system repeatedly interacts with fresh, uncorrelated environments, where each environment is projectively measured before and after the interaction.}\label{fig:Cover}
\end{figure}

In this Letter, we address this gap by deriving universal precision bounds for generic observables in open quantum systems.
We consider a general setup in which a system interacts with its environment at an arbitrary coupling strength, and two-point measurements are performed on the total system, yielding stochastic outcomes from which trajectory observables are defined (see Fig.~\ref{fig:Cover}).
To capture dynamical factors relevant to precision, we introduce a novel quantity, termed the \emph{forward-backward asymmetry}, which quantifies the disparity between forward and backward processes.
Within a thermodynamically consistent framework, we prove that the relative fluctuation of any time-antisymmetric current is jointly constrained by entropy production and the forward-backward asymmetry [Eq.~\eqref{eq:main.res.1}].
This finding reveals the mechanism of precision enhancement in the quantum regime: currents can attain higher precision due to significant asymmetry, even at low dissipation.
For general observables, we further show that their precision is always bounded from below by a generalized activity term that characterizes environmental changes [Eq.~\eqref{eq:main.res.2}].
Notably, this result substantially tightens and extends previous findings to more general setups \cite{Hasegawa.2021.PRL,Hasegawa.2021.PRL2}.
Taken together, these bounds generalize both the TUR and KUR to the strong-coupling quantum regime; remarkably, they are saturable and broadly applicable, and establish fundamental limits on the precision of observables in open quantum systems.
We stress that these results are distinct from metrological bounds \cite{Paris.2009.IJQI,Giovannetti.2011.NP,Montenegro.2025.PR}, which constrain the fluctuations of {\it trajectory-independent} estimators via the quantum Fisher information, a quantity rooted in the geometry of quantum state distinguishability \cite{fnt2}.

\sectionprl{Setup}We consider a finite-dimensional system $S$, which sequentially interacts with fresh, uncorrelated environments $E$ over a total duration of $T$ \cite{Manzano.2018.PRX}. 
During each interaction, the composite system undergoes unitary evolution governed by the total Hamiltonian
\begin{equation}\label{eq:tot.Ham}
	H=H_S+H_E+H_I,
\end{equation}
where $H_S$, $H_E$, and $H_I$ denote the system, environment, and interaction Hamiltonians, respectively.
The coupling between the system and the environment can be arbitrarily strong.
Two-point measurements are performed on both the system and the environment.
The system is projectively measured only at the initial time $t=0$ and the final time $t=T$, using an orthonormal basis $\{\ket{n}\}$, with outcomes $n$ and $m$, respectively.
The post-measurement system state is $\varrho_S = \sum_n p_n \dyad{n}$, where $\{p_n\}$ is the probability distribution over measurement outcomes.
During each time interval $[(i-1)\tau,i\tau]$, for $i=1,2,\dots,N$, the system is coupled to a new, uncorrelated environment initialized in a generic state $\varrho_{E}=\sum_\mu p_\mu\dyad{\mu}$.
The environment is measured before and after each interaction in the basis $\{\ket{\mu}\}$, yielding outcomes $(\nu_i,\mu_i)$.
Note that the initial projective measurement on the environment does not disturb its state.
A stochastic trajectory of measurement outcomes is denoted by $\gamma=\{n,(\nu_1,\mu_1),\dots,(\nu_N,\mu_N),m\}$.
For $N=1$, scenario (b) reduces to scenario (a) in Fig.~\ref{fig:Cover}.
As shown later, this setup establishes a thermodynamically consistent framework that captures non-Markovian dynamics of finite-dimensional systems with arbitrarily engineered system-environment couplings; it also provides operational access to observables and recovers Markovian dynamics in the limit $\tau \to 0$ \cite{fnt3}.
The reduced dynamics of the system for each interaction is described by a completely positive trace-preserving map
\begin{equation}
	\mca{E}(\circ)\coloneqq\sum_{\mu,\nu}M_{\mu\nu}(\circ)M_{\mu\nu}^\dagger,
\end{equation}
where the Kraus operators are defined as $M_{\mu\nu}=\sqrt{p_\nu}\mel{\mu}{U}{\nu}$, with $U$ denoting the unitary operator generated by the total Hamiltonian $H$ in Eq.~\eqref{eq:tot.Ham}.
These operators satisfy the completeness condition $\sum_{\mu,\nu}M_{\mu\nu}^\dagger M_{\mu\nu}=\mds{1}_S$.
The probability of observing the stochastic trajectory $\gamma$ is given by
\begin{equation}
	\mds{P}(\gamma)=p_n|\mel{m}{M_{\mu_N\nu_N}\dots M_{\mu_1\nu_1}}{n}|^2.
\end{equation}
We are interested in trajectory-dependent observables $\phi(\gamma)$, which include any time-integrated counting observables.
Our goal is to elucidate the relationship between the relative fluctuation $\mvar[\phi]/\ev{\phi}^2$ and thermodynamic costs. 
Here, $\mvar[\phi]\coloneqq\ev{\phi^2}-\ev{\phi}^2$ is the observable variance, and $\ev{\circ}$ denotes the ensemble average over all stochastic trajectories.

Next, we briefly describe the thermodynamics of open quantum systems and introduce several key quantities.
To ensure thermodynamical consistency, we assume that the initial state of each environment is a thermal Gibbs state at inverse temperature $\beta$ (i.e., $\varrho_{E}=e^{-\beta H_E}/\tr e^{-\beta H_E}$) \cite{fnt1}.
Within the framework of quantum thermodynamics, entropy production---quantifying the degree of thermodynamic irreversibility---is defined as the sum of the von Neumann entropy change $\Delta S$ in the system and the heat $Q$ dissipated into the environment \cite{Esposito.2010.NJP}:
\begin{equation}
	\Sigma=\Delta S+\beta Q.
\end{equation}
For simplicity, we assume that the system is initially in a stationary state, i.e., $\varrho_S = \mca{E}(\varrho_S)$, as the generalization to arbitrary initial states is straightforward.
Under this assumption, the system entropy change $\Delta S$ vanishes, and the entropy production reduces to heat dissipation and takes the explicit form: $\Sigma=ND(\varrho_{SE}'\|\varrho_S\otimes\varrho_E)$ \cite{Supp.PhysRev}.
Here, $\varrho_{SE}'=U(\varrho_S\otimes\varrho_E)U^\dagger$ represents the ensemble state of the composite system immediately after each interaction, and $D(\cdot\|\cdot)$ denotes the quantum relative entropy.

The thermodynamics of open quantum systems can also be formulated at the level of individual trajectories.
To proceed, we define the time-reversed (backward) process using an antiunitary time-reversal operator $\Theta=\Theta_S\otimes\Theta_E$, which satisfies $\Theta i=-i\Theta$ and $\Theta\Theta^\dagger=\Theta^\dagger\Theta=\mds{1}$ \cite{Campisi.2011.RMP}.
The initial states of the system and environment in the backward process are $\Theta_S\varrho_S\Theta_S^\dagger$ and $\Theta_E\varrho_E\Theta_E^\dagger$, respectively, and projective measurements are performed in the time-reversed bases $\{\Theta_S\ket{n}\}$ and $\{\Theta_E\ket{\mu}\}$.
The unitary operator $\widetilde{U}$ governing the evolution in the backward process is generated by the time-reversed Hamiltonian $\Theta H\Theta^\dagger$.
The probability of observing the time-reversed trajectory $\widetilde{\gamma}=\{m,(\mu_N,\nu_N),\dots,(\mu_1,\nu_1),n\}$ in the backward process is given by
\begin{equation}
	\widetilde{\mds{P}}(\widetilde{\gamma})=p_m|\mel{n}{\Theta_S^\dagger\widetilde{M}_{\nu_1\mu_1}\dots\widetilde{M}_{\nu_N\mu_N}\Theta_S}{m}|^2,
\end{equation}
where the Kraus operators in the backward process are defined as $\widetilde{M}_{\nu\mu}=\sqrt{p_\mu}\mel{\nu}{\Theta_E^\dagger\widetilde{U}\Theta_E}{\mu}$.
The entropy production $\Sigma$ can then be expressed as the average logarithmic ratio between forward and backward trajectory probabilities:
\begin{equation}
	\Sigma=\ev{\ln\frac{\mds{P}(\gamma)}{\widetilde{\mds{P}}(\widetilde{\gamma})}}.
\end{equation}
This representation highlights $\Sigma$ as a measure of time-reversal symmetry breaking due to thermodynamic dissipation.
In addition to $\Sigma$, another key quantity relevant to time-reversal symmetry is the relative entropy between the forward and backward probabilities of the \emph{same} trajectory, defined as
\begin{equation}
	\Sigma_*\coloneqq\ev{\ln\frac{\mds{P}(\gamma)}{\widetilde{\mds{P}}(\gamma)}}\ge 0.
\end{equation}
The quantity $\Sigma_*$ is strictly positive whenever the forward and backward path probabilities, $\mds{P}(\gamma)$ and $\widetilde{\mds{P}}(\gamma)$, differ, and it vanishes only when the two processes are identical (i.e., $\widetilde{\mds{P}} \equiv \mds{P}$). In the presence of magnetic fields (i.e., odd variables under time reversal), the asymmetry $\mds{P} \nequiv \widetilde{\mds{P}}$ generally holds, leading to a nonzero $\Sigma_*$ (e.g., Sec.~S5 of Ref.~\cite{Supp.PhysRev}). Even without magnetic fields, Markovian dynamics that preserve quantum coherence in the system's state can also give rise to a positive asymmetry, as discussed later. More generally, we can prove that the ability of the dynamics to generate quantum entanglement between the system and its environment is a necessary condition for $\Sigma_*$ to be nonzero \cite{Supp.PhysRev}. Additionally, we show that $\Sigma_*$ vanishes in relevant cases, including incoherent Markovian dynamics and a class of thermal operations \cite{Supp.PhysRev}. Thus, $\Sigma_*$ quantifies the asymmetry between forward and backward processes, arising from dynamical features such as quantum coherence, entanglement, or external magnetic fields. Owing to both its definition and its physical interpretation, we refer to $\Sigma_*$ as the {\it forward-backward asymmetry}.
Together, $\Sigma$ and $\Sigma_*$ provide complementary characterizations of time-reversal symmetry breaking in open quantum systems: the former captures thermodynamic irreversibility, while the latter reflects intrinsic dynamical asymmetries.

\sectionprl{Main results}With the key quantities defined above, we are now ready to present our main results; the proof is deferred to Appendix A.
We begin by considering current-type observables that satisfy the time-antisymmetry condition $\phi(\widetilde{\gamma})=-\phi(\gamma)$.
These include, but are not limited to, time-integrated currents commonly studied in conventional TUR formulations \cite{Gingrich.2016.PRL}.
As our first main result, we prove that the relative fluctuation of any time-antisymmetric current is bounded from below by a function of both the entropy production $\Sigma$ and the forward-backward asymmetry $\Sigma_*$:
\begin{equation}\label{eq:main.res.1}
	\frac{\mvar[\phi]}{\ev{\phi}^2}\ge f\qty(\Sigma+\Sigma_*).
\end{equation}
Here, $f(x)= 4[\Phi(x/2)/x]^2-1\in(0,+\infty)$ is a monotonically decreasing function, and $\Phi$ denotes the inverse function of $x\tanh(x)$.
The relation \eqref{eq:main.res.1} can be interpreted as a generalized quantum TUR.
It implies that achieving high precision in current-type observables necessarily requires either high dissipation or significant forward-backward asymmetry.
In other words, both thermodynamic irreversibility and dynamical asymmetry contribute to constraining fluctuations in open quantum systems.
A generalization of this result to arbitrary initial states is provided in Ref.~\cite{Supp.PhysRev}, where a boundary term arising from initial conditions is incorporated into the bound.

Several remarks on the result \eqref{eq:main.res.1} are in order.
First, the lower-bound function behaves as $f(x)\approx 2/x$ for $x\ll 1$, and decays exponentially to zero as $x\to\infty$.
This reflects the fact that the bound is applicable not only to time-extensive but also to time-intensive currents.
Second, the result highlights that the asymmetry $\Sigma_*$ plays a role equally important to that of the entropy production $\Sigma$ in constraining current precision. 
As demonstrated numerically later, the relative fluctuation of currents cannot, in general, be bounded solely by entropy production.
We note that contributions to the asymmetry, such as magnetic fields and initial conditions, have classical counterparts, whereas genuinely quantum features like coherence and entanglement do not (cf.~Eq.~(S102) in Ref.~\cite{Supp.PhysRev}).
Third, in the case where $\Sigma_*=0$ (i.e., the forward and backward processes coincide), the bound \eqref{eq:main.res.1} reduces to a TUR previously derived from the detailed fluctuation theorem \cite{Hasegawa.2019.PRL,Timpanaro.2019.PRL,Vu.2020.JPA}. 
This fluctuation-theorem TUR is known to be tight and saturable in certain cases, indicating that our generalized result inherits this desirable property.
It is worth noting that another TUR, derived from the fluctuation theorem, also holds in our setup \cite{Potts.2019.PRE}. However, this relation involves the statistics of currents in the {\it backward} process, thus rendering it unsuitable for characterizing the ultimate precision limits of currents in the {\it forward} dynamics.
Finally, Markovian dynamics can be recovered from our setup in the limit of short-time interactions (i.e., $\tau \to 0$) \cite{Ciccarello.2022.PR}. For Markovian processes governed by a Hamiltonian $H$ and jump operators $\{L_k\}_k$ that satisfy the local detailed balance condition \cite{Horowitz.2013.NJP}, the forward-backward asymmetry $\Sigma_*$ can be lower bounded in the short-time limit as \cite{Supp.PhysRev}
\begin{equation}
	\Sigma_*\ge\frac{8}{9}\frac{|\ev{[H,\msf{L}]}_S|^2}{\ev{2H^2+\msf{L}^2/2}_S}T^2,
\end{equation}
where $\msf{L}\coloneqq\sum_kL_k^\dagger L_k$ and $\ev{\circ}_S\coloneqq\tr(\circ\varrho_S)$.
This inequality confirms that $\Sigma_*>0$ whenever $H$ and $\msf{L}$ do not commute, a clear signature of quantum coherent dynamics.
A general lower bound of $\Sigma_*$ for finite times is presented in Ref.~\cite{Supp.PhysRev}.

Next, we turn to the case of generic observables, without imposing any time-antisymmetry condition.
Moreover, both the system and the environment may be initialized in arbitrary states, which need not be stationary or thermal Gibbs states, and projective measurements can be taken in any basis.
Let $\mca{I}$ denote the subset of stochastic trajectories in which no change is detected in the environment, $\mca{I}=\{\gamma\,|\,\mu_i=\nu_i\,\forall i\}$.
We impose a minimal condition on observables that $\phi(\gamma)=0$ for any $\gamma\in\mca{I}$.
This covers a wide range of observables, such as energy currents, particle transport, and the total number of quantum jumps detected in the environment.
For this general setup, we prove the following lower bound on the relative fluctuation of any such observable:
\begin{equation}\label{eq:main.res.2}
	\frac{\mvar[\phi]}{\ev{\phi}^2}\ge\frac{1}{\msc{P}^{-1}-1}.
\end{equation}
Here, $\msc{P}$ denotes the probability of detecting no environmental change, defined as $\msc{P}\coloneqq\sum_{\gamma\in\mca{I}}\mds{P}(\gamma)<1$.
This constitutes our second main result, which holds under highly general conditions: for arbitrary initial states and system-environment interactions.
Since $\msc{P}$ quantifies the inactivity of the environment, its inverse $\msc{P}^{-1}$ can be interpreted as an activity term.
The bound \eqref{eq:main.res.2} thus implies that achieving high precision requires correspondingly high activity.
In the short-time limit of Markovian dynamics, the denominator $\msc{P}^{-1}-1$ reduces exactly to the dynamical activity \cite{Supp.PhysRev}.
Therefore, the inequality \eqref{eq:main.res.2} can be regarded as a generalized quantum KUR. 

We provide several remarks on the result \eqref{eq:main.res.2}.
First, the inequality is tight and can be saturated.
Specifically, the equality holds for the observable $\phi$ that simply detects any environmental change, i.e., $\phi(\gamma)=1$ for all $\gamma\notin\mca{I}$ and zero otherwise.
Second, although we focus on the case where $\mca{I}$ represents no-jump trajectories, the result remains valid for any subset $\mca{I}$ of trajectories, as long as the condition $\phi(\gamma)=0$ for any $\gamma\in\mca{I}$ is fulfilled.
This generality significantly broadens the applicability of the bound across a wide range of physical settings.
Third, the inequality \eqref{eq:main.res.2} generalizes and sharpens a pivotal result previously reported in Ref.~\cite{Hasegawa.2021.PRL}.
In particular, when the environment is initialized in the pure state $\dyad{0}$, it was shown that the precision of observables satisfies $\mvar[\phi]/\ev{\phi}^2\ge 1/(\msc{A}-1)$, where $\msc{A}\coloneqq\tr{(V_0^\dagger V_0)^{-1}\varrho_S}$ is known as the survival activity \cite{Hasegawa.2021.PRL} and $V_0=M_{00}^N$.
Using the inequality $|\tr(AB)|^2\le\tr(A^\dagger A)\tr(B^\dagger B)$ and noting that $\msc{P}=\tr{(V_0^\dagger V_0)\varrho_S}$, we obtain
\begin{equation}
	1\le \tr{(V_0^\dagger V_0)^{-1}\varrho_S}\tr{(V_0^\dagger V_0)\varrho_S}=\msc{A}\msc{P},
\end{equation}
which immediately yields $\msc{P}^{-1}\le\msc{A}$.
Therefore, the bound \eqref{eq:main.res.2} is strictly tighter than the previous finding, and more importantly, it naturally extends to arbitrary initial states of the environment.
Finally, in the case of Markovian dynamics with arbitrary Hamiltonian and jump operators, the inactivity $\msc{P}$ can be explicitly expressed as $\msc{P}=\tr(e^{-iH_{\rm eff}T}\varrho_Se^{iH_{\rm eff}^\dagger T})$, where $H_{\rm eff}= H-(i/2)\sum_kL_k^\dagger L_k$ is the effective non-Hermitian Hamiltonian that describes no-jump evolution.
In this setting, Ref.~\cite{Hasegawa.2021.PRL2} derived an alternative precision bound using the Loschmidt echo approach: $\mvar[\phi]/\ev{\phi}^2\ge 1/(\eta^{-1}-1)$, where $\eta=|\tr(e^{-iH_{\rm eff}T}\varrho_S)|^2$ is the Loschmidt echo.
Applying the Cauchy-Schwarz inequality gives
\begin{equation}
	\eta\le\tr(e^{-iH_{\rm eff}T}\varrho_Se^{iH_{\rm eff}^\dagger T})\tr(\varrho_S)=\msc{P},
\end{equation}
which implies that the result \eqref{eq:main.res.2} is always sharper than the Loschmidt echo-based bound in this setting.

\begin{figure*}[t]
\centering
\includegraphics[width=1\linewidth]{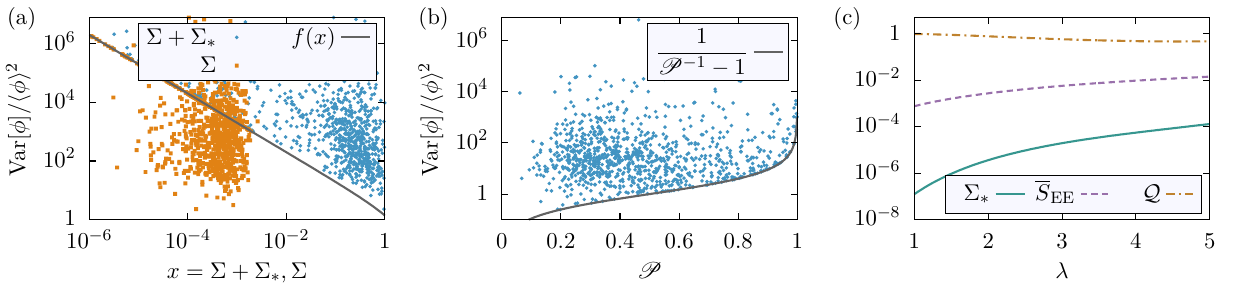}
\protect\caption{Numerical illustration of the main results \eqref{eq:main.res.1} and \eqref{eq:main.res.2} for a qubit interacting with a finite-dimensional environment. (a) Blue circles and orange squares represent the relative fluctuation of current-type observables plotted against $\Sigma+\Sigma_*$ and $\Sigma$, respectively. The solid line depicts the lower-bound function $f(x)$. (b) Blue circles represent the relative fluctuation of generic observables plotted against $\msc{P}$. The solid line shows the lower bound $1/(\msc{P}^{-1}-1)$. (c) Solid, dashed, dash-dotted lines represent the asymmetry $\Sigma_*$, the entanglement entropy $\overline{S}_{\rm EE}$, and the quality factor $\mca{Q}$, respectively, as functions of the coupling strength $\lambda$. Parameters are chosen as follows: the environment dimension $d_E$ is a random integer in the range $[2,5]$; the eigenvalues of the environmental Hamiltonian $H_E$ are sampled from $[0,0.1]$; $V_S$ and $V_E$ are random Hermitian operators with matrix elements in $[-1-i,1+i]$; and fixed parameters are $\omega_z=1$, $\omega_x=0.1$, $\lambda=5$, $\beta=1$, and $T=5$.}\label{fig:Qubit}
\end{figure*}

\sectionprl{Example}We illustrate our results [Eqs.~\eqref{eq:main.res.1} and \eqref{eq:main.res.2}] in a qubit system interacting with a finite-dimensional environment.
The total Hamiltonian is given by
\begin{equation}
	H=\frac{1}{2}(\omega_z\sigma_z+\omega_x\sigma_x)\otimes\mds{1}_E+\mds{1}_S\otimes H_E+\lambda V_S\otimes V_E,
\end{equation}
where $\sigma_{x,z}$ are the Pauli matrices, and $\lambda$ denotes the coupling strength between the qubit and the environment.
The environment may represent a heat bath or an ancillary qudit.
Two-point measurements are performed on the total system only at the initial and final times (i.e., $N=1$).
In the strong-coupling regime (i.e., large $\lambda$), the system and the environment become significantly entangled, leading to the potential for precision enhancement and violation of conventional TURs.

To demonstrate the bounds, we fix the energy eigenstates of the environmental Hamiltonian $H_E$, while randomly sampling its eigenvalues, as well as the matrix elements of the Hermitian operators $V_S$ and $V_E$.
Observables are defined as $\phi(\gamma)=c_{\mu\nu}$, where $\{\nu,\mu\}$ are the measurement outcomes on the environment at the initial and final times, and the coefficients $c_{\mu\nu}$ are also randomly chosen.
For each instance of the parameter set, we compute the relative fluctuation of the observable, the entropy production $\Sigma$, the forward-backward asymmetry $\Sigma_*$, and the inactivity $\msc{P}$.
The numerical results are summarized in Fig.~\ref{fig:Qubit}.

For current-type observables that satisfy $c_{\mu\nu} = -c_{\nu\mu}$, Fig.~\ref{fig:Qubit}(a) shows that their relative fluctuation is always bounded from below by $f(\Sigma + \Sigma_*)$.
In contrast, significant violations occur when the forward-backward asymmetry $\Sigma_*$ is ignored, demonstrating that fluctuations cannot, in general, be constrained solely by entropy production.
For generic observables, Fig.~\ref{fig:Qubit}(b) confirms that their precision is always bounded from below by the inverse of the generalized activity term, $1/(\msc{P}^{-1}-1)$, in agreement with Eq.~\eqref{eq:main.res.2}.
Both bounds are shown to be tight and saturable across the sampled parameter space.

We next investigate the relationship between the forward-backward asymmetry $\Sigma_*$ and the quantum entanglement generated between the system and the environment.
Since the composite system is initially in a pure product state after the projective measurements and subsequently evolves unitarily, quantum entanglement can be quantified via the entanglement entropy.
Specifically, we define the average entanglement entropy as
\begin{equation}
	\overline{S}_{\rm EE}\coloneqq\sum_{n,\nu}p_np_\nu S_{\rm EE}(\varrho_{n,\nu}'),
\end{equation}
where $\varrho_{n,\nu}'=\tr_E(U\dyad{n,\nu}U^\dagger)$ is the conditional system state at the final time and $S_{\rm EE}(\varrho)=-\tr(\varrho\ln\varrho)$ is the von Neumann entropy.
We vary the coupling strength $\lambda$ while keeping other parameters fixed, and plot $\Sigma_*$, $\overline{S}_{\rm EE}$, and the quality factor $\mca{Q}\coloneqq(\mvar[\phi]/\ev{\phi}^2)/f(\Sigma)$ as functions of $\lambda$ in Fig.~\ref{fig:Qubit}(c). Note that $\mca{Q}<1$ indicates the violation of the conventional TUR \cite{Hasegawa.2019.PRL,Timpanaro.2019.PRL,Vu.2020.JPA} based solely on entropy production. As shown, both the forward-backward asymmetry $\Sigma_*$ and the entanglement entropy $\overline{S}_{\rm EE}$ monotonically increase with $\lambda$, while the quality factor $\mca{Q}$ significantly drops below $1$. This supports that quantum entanglement contributes to the emergence of dynamical asymmetry and enables precision enhancement beyond the classical TUR.

\sectionprl{Conclusion and outlook}In this Letter, we established universal precision bounds for both time-antisymmetric (current-type) and generic observables in general open quantum systems subjected to two-point measurement protocols.
Our analysis reveals that, beyond the well-known role of dissipation, the forward-backward asymmetry serves as a fundamental factor limiting the precision of currents.
We further anticipate that this notion of asymmetry will play an essential role in nonequilibrium thermodynamics more broadly, as evidenced in Ref.~\cite{Supp.PhysRev}, where it is shown to govern a generalized TUR for underdamped Langevin systems.
Meanwhile, for general observables, we showed that the generalized activity, which captures the underlying kinetic structure of quantum trajectories, imposes a tight constraint on their precision.
These results offer a comprehensive and experimentally relevant framework for understanding the fundamental limits of precision in open quantum systems.

We conclude by highlighting several open directions for future research. 
A first important direction is to extend our framework to more realistic settings, including continuous-variable systems \cite{Braunstein.2005.RMP} and infinitely large environments \cite{Vega.2017.RMP}, which are fundamental to quantum optics, condensed matter, and many-body physics. Establishing precision bounds in such regimes would clarify the universality of thermodynamic constraints on fluctuations.
Second, it remains open whether structurally refined precision-cost constraints emerge for widely studied observables, such as counting observables in conventional TUR and KUR formulations, potentially revealing a hierarchy of precision limits.
Finally, extending our results to incorporate active quantum feedback control may provide guiding principles for the design of quantum clocks, sensors, and other precision devices \cite{Parrondo.2015.NP}.

\sectionprl{Note added}We recently became aware that results related to Eq.~\eqref{eq:main.res.2} have been derived in Ref.~\cite{Hasegawa.2025.PRL} using the Petrov inequality and in Ref.~\cite{Hegde.2026.PRE} using a fluctuation theorem approach.

\begin{acknowledgments}
\sectionprl{Acknowledgments}{T.V.V.} was supported by JSPS KAKENHI Grants No.~JP23K13032 and No.~JP 26K00022. {K.S.} was supported by JSPS KAKENHI Grant No.~JP23K25796 and No.~JP26H02015.
\end{acknowledgments}

\sectionprl{Data availability}The data are not publicly available upon publication. The data are available from the authors upon reasonable request.

\onecolumngrid
\begin{center}
    \textbf{End Matter}
\end{center}
\twocolumngrid

\sectionprl{Appendix A: Proof of Eqs.~\eqref{eq:main.res.1} and \eqref{eq:main.res.2}}For any observable $\phi$ that satisfies the time-antisymmetry condition, its first and second moments can be expressed as
\begin{align}
	\ev{\phi}&=\frac{1}{2}\sum_{\gamma}\phi(\gamma)[\mds{P}(\gamma)-\mds{P}(\widetilde{\gamma})],\\
	\ev{\phi^2}&=\frac{1}{2}\sum_{\gamma}\phi(\gamma)^2[\mds{P}(\gamma)+\mds{P}(\widetilde{\gamma})].
\end{align}
Applying the Cauchy-Schwarz inequality to these expressions yields
\begin{align}
	\ev{\phi}^2&\le\frac{1}{4}\sum_{\gamma}\frac{[\mds{P}(\gamma)-\mds{P}(\widetilde{\gamma})]^2}{\mds{P}(\gamma)+\mds{P}(\widetilde{\gamma})}\sum_{\gamma}\phi(\gamma)^2[\mds{P}(\gamma)+\mds{P}(\widetilde{\gamma})]\notag\\
	&=\ell\ev{\phi^2},\label{eq:proof.tmp1}
\end{align}
where we define 
\begin{equation}
	\ell\coloneqq\frac{1}{2}\sum_{\gamma}\frac{[\mds{P}(\gamma)-\mds{P}(\widetilde{\gamma})]^2}{\mds{P}(\gamma)+\mds{P}(\widetilde{\gamma})}.
\end{equation}
From Eq.~\eqref{eq:proof.tmp1}, we obtain the following lower bound on the relative fluctuation of the observable $\phi$:
\begin{equation}\label{eq:proof.tmp2}
	\frac{\mvar[\phi]}{\ev{\phi}^2}\ge\ell^{-1}-1.
\end{equation}
We only need to upper bound $\ell$ in terms of the entropy production $\Sigma$ and the forward-backward asymmetry $\Sigma_*$.
To this end, note that the entropy production can be calculated as
\begin{align}
	\Sigma&=\sum_{\gamma}\mds{P}(\gamma)\ln\frac{\mds{P}(\gamma)}{\widetilde{\mds{P}}(\widetilde{\gamma})}\notag\\
	&=\sum_{\gamma}\mds{P}(\gamma)\ln\frac{\mds{P}(\gamma)}{\mds{P}(\widetilde{\gamma})}-\sum_{\gamma}\mds{P}(\gamma)\ln\frac{\widetilde{\mds{P}}(\widetilde{\gamma})}{\mds{P}(\widetilde{\gamma})}\notag\\
	&=\sum_{\gamma}\mds{P}(\gamma)\ln\frac{\mds{P}(\gamma)}{\mds{P}(\widetilde{\gamma})}-\sum_{\gamma}\mds{P}(\gamma)\ln\frac{\mds{P}(\gamma)}{\widetilde{\mds{P}}(\gamma)}\notag\\
	&=\sum_{\gamma}\mds{P}(\gamma)\ln\frac{\mds{P}(\gamma)}{\mds{P}(\widetilde{\gamma})}-\Sigma_*.
\end{align}
Here, we use the equality $\widetilde{\mds{P}}(\widetilde{\gamma})/\mds{P}(\widetilde{\gamma})=\mds{P}(\gamma)/\widetilde{\mds{P}}(\gamma)$ (see Appendix B) to get the third line.
Therefore, we obtain
\begin{align}
	\Sigma+\Sigma_*&=\sum_{\gamma}\mds{P}(\gamma)\ln\frac{\mds{P}(\gamma)}{\mds{P}(\widetilde{\gamma})}\notag\\
	&=\frac{1}{2}\sum_{\gamma}[\mds{P}(\gamma)-\mds{P}(\widetilde{\gamma})]\ln\frac{\mds{P}(\gamma)}{\mds{P}(\widetilde{\gamma})}.
\end{align}
To proceed, we define
\begin{align}
	\sigma(\gamma)&\coloneqq [\mds{P}(\gamma)-\mds{P}(\widetilde{\gamma})]\ln\frac{\mds{P}(\gamma)}{\mds{P}(\widetilde{\gamma})},\\
	a(\gamma)&\coloneqq \mds{P}(\gamma)+\mds{P}(\widetilde{\gamma}).
\end{align}
Evidently, these quantities satisfy the relations $\sum_{\gamma}\sigma(\gamma)=2(\Sigma+\Sigma_*)$ and $\sum_{\gamma}a(\gamma)=2$.
We now derive an upper bound on $\ell$ in terms of $\Sigma+\Sigma_*$.
By algebraic manipulations and an application of Jensen's inequality, we can show that
\begin{align}
	\ell&=\frac{1}{2}\sum_{\gamma}\frac{[\mds{P}(\gamma)-\mds{P}(\widetilde{\gamma})]^2}{\mds{P}(\gamma)+\mds{P}(\widetilde{\gamma})}\notag\\
	&=\frac{1}{2}\sum_{\gamma}\frac{\sigma(\gamma)^2}{4a(\gamma)}\Phi\qty[\frac{\sigma(\gamma)}{2a(\gamma)}]^{-2}\notag\\
	&\le\frac{1}{4}(\Sigma+\Sigma_*)^2\Phi\qty(\frac{\Sigma+\Sigma_*}{2})^{-2}.\label{eq:proof.tmp3}
\end{align}
Here, we exploit the convexity of $(x^2/y)\Phi(x/2y)^{-2}$ to obtain the last line.
Combining Eqs.~\eqref{eq:proof.tmp2} and \eqref{eq:proof.tmp3} leads to the desired relation \eqref{eq:main.res.1}:
\begin{equation}
	\frac{\mvar[\phi]}{\ev{\phi}^2}\ge f(\Sigma+\Sigma_*).
\end{equation}
According to Proposition 1 in Ref.~\cite{Supp.PhysRev}, the lower-bound function can also be alternatively expressed as
\begin{equation}
	f(\Sigma+\Sigma_*)=\csch^2\qty[\Phi\qty(\frac{\Sigma+\Sigma_*}{2})]\ge\frac{2}{e^{\Sigma+\Sigma_*}-1}.
\end{equation}

Next, we prove the relation \eqref{eq:main.res.2}.
Since $\phi(\gamma)=0$ for any $\gamma\in\mca{I}$, the first and second moments of generic observable $\phi$ can be calculated as
\begin{align}
	\ev{\phi}&=\sum_{\gamma\notin\mca{I}}\phi(\gamma)\mds{P}(\gamma),\\
	\ev{\phi^2}&=\sum_{\gamma\notin\mca{I}}\phi(\gamma)^2\mds{P}(\gamma).
\end{align}
Applying the Cauchy-Schwarz inequality, the first moment can be upper bounded by the second moment as
\begin{align}
	\ev{\phi}^2&=\qty[\sum_{\gamma\notin\mca{I}}\phi(\gamma)\mds{P}(\gamma)]^2\notag\\
	&\le\qty[\sum_{\gamma\notin\mca{I}}\mds{P}(\gamma)]\qty[\sum_{\gamma\notin\mca{I}}\phi(\gamma)^2\mds{P}(\gamma)]\notag\\
	&=(1-\msc{P})\ev{\phi^2}.\label{eq:proof.tmp4}
\end{align}
By transforming Eq.~\eqref{eq:proof.tmp4}, the bound \eqref{eq:main.res.2} can be readily obtained.

\sectionprl{Appendix B: Proof of $\widetilde{\mds{P}}(\widetilde{\gamma})/\mds{P}(\widetilde{\gamma})=\mds{P}(\gamma)/\widetilde{\mds{P}}(\gamma)$}Noting that $\widetilde{\gamma}=\{m,(\mu_N,\nu_N),\dots,(\mu_1,\nu_1),n\}$, the path probabilities can be explicitly expressed as follows:
\begin{align}
	\mds{P}(\gamma)&=p_n|\mel{m}{M_{\mu_N\nu_N}\dots M_{\mu_1\nu_1}}{n}|^2,\\
	\mds{P}(\widetilde{\gamma})&=p_m|\mel{n}{M_{\nu_1\mu_1}\dots M_{\nu_N\mu_N}}{m}|^2,\\
	\widetilde{\mds{P}}(\widetilde{\gamma})&=p_m|\mel{n}{\Theta_S^\dagger\widetilde{M}_{\nu_1\mu_1}\dots\widetilde{M}_{\nu_N\mu_N}\Theta_S}{m}|^2,\\
	\widetilde{\mds{P}}(\gamma)&=p_n|\mel{m}{\Theta_S^\dagger\widetilde{M}_{\mu_N\nu_N}\dots\widetilde{M}_{\mu_1\nu_1}\Theta_S}{n}|^2.
\end{align}
In addition, the Kraus operators in the backward process are related to those in the forward process as
\begin{align}
	\Theta_S^\dagger\widetilde{M}_{\mu\nu}\Theta_S&=\sqrt{p_\nu}\Theta_S^\dagger\mel{\mu}{\Theta_E^\dagger\widetilde{U}\Theta_E}{\nu}\Theta_S\notag\\
	&=\sqrt{p_\nu}\mel{\mu}{U^\dagger}{\nu}\notag\\
	&=\sqrt{p_\nu/p_\mu}M_{\nu\mu}^\dagger.
\end{align}
Here, we use the fact $\widetilde{U}=\Theta U^\dagger \Theta^\dagger$ to obtain the second line.
Using this relation, the path probabilities can be simplified further as follows:
\begin{align}
	\widetilde{\mds{P}}(\widetilde{\gamma})&=p_m|\mel{n}{\Theta_S^\dagger\widetilde{M}_{\nu_1\mu_1}\dots\widetilde{M}_{\nu_N\mu_N}\Theta_S}{m}|^2\notag\\
	&=p_m\frac{p_{\mu_1}\dots p_{\mu_N}}{p_{\nu_1}\dots p_{\nu_N}}|\mel{n}{M_{\mu_1\nu_1}^\dagger\dots M_{\mu_N\nu_N}^\dagger}{m}|^2\notag\\
	&=p_m\frac{p_{\mu_1}\dots p_{\mu_N}}{p_{\nu_1}\dots p_{\nu_N}}|\mel{m}{M_{\mu_N\nu_N}\dots M_{\mu_1\nu_1}}{n}|^2\notag\\
	&=\frac{p_m}{p_n}\frac{p_{\mu_1}\dots p_{\mu_N}}{p_{\nu_1}\dots p_{\nu_N}}\mds{P}(\gamma),\\
	\widetilde{\mds{P}}(\gamma)&=p_n|\mel{m}{\Theta_S^\dagger\widetilde{M}_{\mu_N\nu_N}\dots\widetilde{M}_{\mu_1\nu_1}\Theta_S}{n}|^2\notag\\
	&=p_n\frac{p_{\nu_1}\dots p_{\nu_N}}{p_{\mu_1}\dots p_{\mu_N}}|\mel{m}{M_{\nu_N\mu_N}^\dagger\dots M_{\nu_1\mu_1}^\dagger}{n}|^2\notag\\
	&=p_n\frac{p_{\nu_1}\dots p_{\nu_N}}{p_{\mu_1}\dots p_{\mu_N}}|\mel{n}{M_{\nu_1\mu_1}\dots M_{\nu_N\mu_N}}{m}|^2\notag\\
	&=\frac{p_n}{p_m}\frac{p_{\nu_1}\dots p_{\nu_N}}{p_{\mu_1}\dots p_{\mu_N}}\mds{P}(\widetilde{\gamma}).
\end{align}
Therefore, we can verify that $\widetilde{\mds{P}}(\widetilde{\gamma})\widetilde{\mds{P}}(\gamma)=\mds{P}(\gamma)\mds{P}(\widetilde{\gamma})$, which immediately derives $\widetilde{\mds{P}}(\widetilde{\gamma})/\mds{P}(\widetilde{\gamma})=\mds{P}(\gamma)/\widetilde{\mds{P}}(\gamma)$.


\begin{thebibliography}{81}
\bibitem [{fnt({\natexlab{b}})}]{fnt3}%
  \BibitemOpen
  \href@noop {} {}We note that scenario (a) can reproduce any completely positive trace-preserving map via the Stinespring dilation theorem, provided that the environment is initialized in a pure state. More generally, it can capture a wide range of non-Markovian dynamics, as long as both the system and environment are finite-dimensional.\BibitemShut {Stop}%
\bibitem [{fnt({\natexlab{c}})}]{fnt1}%
  \BibitemOpen
  \href@noop {} {}While we assume that the system interacts with a single thermal bath during each interaction for simplicity, the formulation extends straightforwardly to multiple heat baths at different temperatures.\BibitemShut {Stop}%
\end{thebibliography}
\end{document}


\title{Supplemental Material for \\``Universal Precision Limits in General Open Quantum Systems''}

\author{Tan Van Vu}
\email{tan.vu@yukawa.kyoto-u.ac.jp}
\affiliation{Center for Gravitational Physics and Quantum Information, Yukawa Institute for Theoretical Physics, Kyoto University, Kitashirakawa Oiwakecho, Sakyo-ku, Kyoto 606-8502, Japan}

\author{Ryotaro Honma}
\affiliation{Center for Gravitational Physics and Quantum Information, Yukawa Institute for Theoretical Physics, Kyoto University, Kitashirakawa Oiwakecho, Sakyo-ku, Kyoto 606-8502, Japan}

\author{Keiji Saito}
\affiliation{Department of Physics, Kyoto University, Kyoto 606-8502, Japan}

\begin{abstract}
This Supplemental Material includes detailed analytical calculations, generalized uncertainty relations for Markovian dynamics, the generalization of the result (\FirRes) to arbitrary initial states, and a generalized thermodynamic uncertainty relation for general underdamped Langevin dynamics. 
The equations and figure numbers are prefixed with S [e.g., Eq.~(S1) or Fig.~S1]. 
The numbers without this prefix [e.g., Eq.~(1) or Fig.~1] refer to the items in the main text.
\end{abstract}

\pacs{}
\maketitle

\tableofcontents

\section{Expression of the entropy production $\Sigma$}
The average amount of heat dissipated into the environment during the $i$th interaction between the system and the environment can be calculated as follows:
\begin{align}
	\sum_{\gamma}\mds{P}(\gamma)(\epsilon_{\mu_i}-\epsilon_{\nu_i})&=\sum_{m,n,\{\mu_j,\nu_j\}_{j=1}^N}p_n|\mel{m}{M_{\mu_N\nu_N}\dots M_{\mu_1\nu_1}}{n}|^2(\epsilon_{\mu_i}-\epsilon_{\nu_i})\notag\\
	&=\sum_{\mu_i,\nu_i}(\epsilon_{\mu_i}-\epsilon_{\nu_i})\tr(M_{\mu_i\nu_i}\varrho_SM_{\mu_i\nu_i}^\dagger)\notag\\
	&=\sum_{\mu_i,\nu_i}p_{\nu_i}\epsilon_{\mu_i}\tr(\mel{\mu_i}{U}{\nu_i}\varrho_S\mel{\nu_i}{U^\dagger}{\mu_i})-\sum_{\mu_i,\nu_i}p_{\nu_i}\epsilon_{\nu_i}\tr(\mel{\mu_i}{U}{\nu_i}\varrho_S\mel{\nu_i}{U^\dagger}{\mu_i})\notag\\
	&=\tr{H_EU(\varrho_S\otimes\varrho_E)U^\dagger}-\tr{U(\varrho_S\otimes H_E\varrho_E)U^\dagger}\notag\\
	&=\tr(H_E\varrho_{SE}')-\tr(H_E\varrho_E),
\end{align}
where $\varrho_{SE}'\coloneqq U(\varrho_S\otimes\varrho_E)U^\dagger$.
Noting that $\varrho_S=\tr_E\varrho_{SE}'$ and $\Delta S=0$, the entropy production $\Sigma$ can be computed as follows:
\begin{align}
	\Sigma&=\Delta S+\beta\Delta Q\notag\\
	&=\beta\ev{\sum_{i=1}^{N}(\epsilon_{\mu_i}-\epsilon_{\nu_i})}\notag\\
	&=N\beta\qty[\tr(H_E\varrho_{SE}')-\tr(H_E\varrho_E)]\notag\\
	&=N\qty[-\tr(\varrho_{SE}'\ln\varrho_E)+\tr(\varrho_E\ln\varrho_E)]\notag\\
	&=N\qty[-\tr(\varrho_{SE}'\ln\varrho_E)-\tr(\varrho_S\ln\varrho_S)+\tr(\varrho_S\ln\varrho_S)+\tr(\varrho_E\ln\varrho_E)]\notag\\
	&=N\qty[-\tr{\varrho_{SE}'\ln(\varrho_S\otimes\varrho_E)}+\tr{(\varrho_S\otimes\varrho_E)\ln(\varrho_S\otimes\varrho_E)}]\notag\\
	&=N\qty[-\tr{\varrho_{SE}'\ln(\varrho_S\otimes\varrho_E)}+\tr(\varrho_{SE}'\ln\varrho_{SE}')]\notag\\
	&=ND(\varrho_{SE}'\|\varrho_S\otimes\varrho_E).
\end{align}
In the stationary state, the total entropy production is thus $N$ times the entropy production that occurs during each interaction.

\section{Properties of the forward-backward asymmetry $\Sigma_*$}
\subsection{Fluctuation theorem for stochastic asymmetry}
The definition of the forward-backward asymmetry $\Sigma_*$ gives rise to a notion of stochastic asymmetry at the trajectory level, defined as
\begin{equation}
	\sigma_*(\gamma)\coloneqq\ln\frac{\mds{P}(\gamma)}{\widetilde{\mds{P}}(\gamma)}.
\end{equation}
Evidently, $\ev{\sigma_*(\gamma)}=\Sigma_*$.
We can show that $\sigma_*(\gamma)$ satisfies an integral fluctuation theorem,
\begin{equation}
	\ev{e^{-\sigma_*(\gamma)}}=\sum_\gamma \mds{P}(\gamma)\frac{\widetilde{\mds{P}}(\gamma)}{\mds{P}(\gamma)}=\sum_\gamma \widetilde{\mds{P}}(\gamma)=1.
\end{equation}

\subsection{Vanishing of $\Sigma_*$ in thermal operations}
Here we demonstrate that $\Sigma_*$ vanishes for a certain class of thermal operations, which are unable to generate quantum coherence.
The unitary transformation $U$ is called a thermal operation if it preserves the total energy (i.e., $[U,H_S+H_E]=0$).
Let $H_S=\sum_n\epsilon_n\dyad{n}$ and $H_E=\sum_\mu\epsilon_\mu\dyad{\mu}$ be the spectral decomposition of the system and environment Hamiltonian, respectively.
Consider thermal operations where $H_S$ and $H_E$ are nondegenerate, and the interaction Hamiltonian is given by the form:
\begin{equation}
	H_I=\sum_{(m,\mu),(n,\nu)}h_{\mu\nu}^{mn}\dyad{m,\mu}{n,\nu}\delta(\epsilon_m+\epsilon_\mu-\epsilon_n-\epsilon_\nu),
\end{equation}
which satisfies $|\mca{S}_{m\mu}|\le 2$ for any $(m,\mu)$.
Here, $\mca{S}_{m\mu}\coloneqq\{(n,\nu)\,|\,\epsilon_n+\epsilon_\nu=\epsilon_m+\epsilon_\mu\}$ denotes the set of energy levels $(n,\nu)$ that have the same total energy with $(m,\mu)$.

In order to prove $\Sigma_*=0$, we need only show that
\begin{equation}\label{eq:prob.equ}
	\widetilde{\mds{P}}(\gamma)=\mds{P}(\gamma).
\end{equation}
We first prove that $\mel{m,\mu}{U}{n,\nu}=0$ if $(n,\nu)\notin\mca{S}_{m\mu}$.
Since $[U,H_S+H_E]=0$, it follows that
\begin{align}
	0&=\mel{m,\mu}{[U,H_S+H_E]}{n,\nu}\notag\\
	&=(\epsilon_n+\epsilon_\nu-\epsilon_m-\epsilon_\mu)\mel{m,\mu}{U}{n,\nu}.
\end{align}
It is thus evident that $\mel{m,\mu}{U}{n,\nu}=0$ if $\epsilon_n+\epsilon_\nu-\epsilon_m-\epsilon_\mu\neq 0$ [or equivalently, $(n,\nu)\notin\mca{S}_{m\mu}$].
Next, we prove that $|\mel{m,\mu}{U}{n,\nu}|=|\mel{n,\nu}{U}{m,\mu}|$ for any $(n,\nu)\in\mca{S}_{m\mu}$.
Since this is trivial for the case $(n,\nu)=(m,\mu)$, we need only consider the case $(n,\nu)\neq(m,\mu)$.
Noting that $U=e^{-iH\tau}=e^{-i(H_S+H_E)\tau}e^{-iH_I\tau}$ and $|\mca{S}_{m\mu}|\le 2$, we can calculate as follows:
\begin{align}
	\mel{m,\mu}{U}{n,\nu}&=e^{-i(\epsilon_m+\epsilon_\mu)\tau}\mel{m,\mu}{e^{-iH_I\tau}}{n,\nu}\notag\\
	&=e^{-i(\epsilon_m+\epsilon_\mu)\tau}\sum_{k=0}^\infty\frac{(-i\tau)^k}{k!}\mel{m,\mu}{H_I^k}{n,\nu}\notag\\
	&=e^{-i(\epsilon_m+\epsilon_\mu)\tau}h_{\mu\nu}^{mn}\sum_{k=1}^\infty\frac{(-i\tau)^k}{k!}\sum_{\gamma_k}(h_{\mu\nu}^{mn}h_{\nu\mu}^{nm})^{N_{21}(\gamma_k)}(h_{\mu\mu}^{mm})^{N_{11}(\gamma_k)}(h_{\nu\nu}^{nn})^{N_{22}(\gamma_k)}.\label{eq:umn.tmp1}
\end{align}
Here, $\gamma_k=(s_0\to s_1\to \dots\to s_{k})$ denotes a path of length $k$ connecting $s_0=(m,\mu)$ and $s_{k}=(n,\nu)$ through a sequence of intermediate nodes $s_i\in\{(m,\mu),(n,\nu)\}$ for $1\le i<k$.
The quantities $N_{11}(\gamma_k)$, $N_{22}(\gamma_k)$, $N_{21}(\gamma_k)$, and $N_{12}(\gamma_k)$ denote the number of transitions $(m,\mu)\to(m,\mu)$, $(n,\nu)\to(n,\nu)$, $(n,\nu)\to(m,\mu)$, and $(m,\mu)\to(n,\nu)$ along the path $\gamma_k$, respectively.
Note that $N_{11}(\gamma_k)+N_{22}(\gamma_k)+2N_{21}(\gamma_k)=k-1$ and $N_{12}(\gamma_k)=N_{21}(\gamma_k)+1$.
For each path $\gamma_k$, we denote its reversed counterpart by $\tilde{\gamma}_k\coloneqq(s_k\to\dots\to s_1\to s_0)$.
It is evident that $N_{11}(\tilde{\gamma}_k)=N_{11}(\gamma_k)$, $N_{22}(\tilde{\gamma}_k)=N_{22}(\gamma_k)$, and $N_{12}(\tilde{\gamma}_k)=N_{21}(\gamma_k)$.
Similarly, we obtain
\begin{align}
	\mel{n,\nu}{U}{m,\mu}&=e^{-i(\epsilon_n+\epsilon_\nu)\tau}h_{\nu\mu}^{nm}\sum_{k=1}^\infty\frac{(-i\tau)^k}{k!}\sum_{\tilde{\gamma}_k}(h_{\mu\nu}^{mn}h_{\nu\mu}^{nm})^{N_{12}(\tilde{\gamma}_k)}(h_{\mu\mu}^{mm})^{N_{11}(\tilde{\gamma}_k)}(h_{\nu\nu}^{nn})^{N_{22}(\tilde{\gamma}_k)}\notag\\
	&=e^{-i(\epsilon_n+\epsilon_\nu)\tau}h_{\nu\mu}^{nm}\sum_{k=1}^\infty\frac{(-i\tau)^k}{k!}\sum_{\gamma_k}(h_{\mu\nu}^{mn}h_{\nu\mu}^{nm})^{N_{21}(\gamma_k)}(h_{\mu\mu}^{mm})^{N_{11}(\gamma_k)}(h_{\nu\nu}^{nn})^{N_{22}(\gamma_k)}.\label{eq:umn.tmp2}
\end{align}
Since $|h_{\mu\nu}^{mn}|=|h_{\nu\mu}^{nm}|$, it follows from Eqs.~\eqref{eq:umn.tmp1} and \eqref{eq:umn.tmp2} that $|\mel{m,\mu}{U}{n,\nu}|=|\mel{n,\nu}{U}{m,\mu}|$.

We are now ready to verify Eq.~\eqref{eq:prob.equ}.
The path probabilities can be expressed as follows:
\begin{align}
	\mds{P}(\gamma)&=p_n|\mel{m}{M_{\mu_N\nu_N}\dots M_{\mu_1\nu_1}}{n}|^2\notag\\
	&=p_np_{\nu_1}\dots p_{\nu_N}|\mel{m}{\mel{\mu_N}{U}{\nu_N}\dots\mel{\mu_1}{U}{\nu_1}}{n}|^2,\label{eq:path.prob.1}\\
	\widetilde{\mds{P}}(\gamma)&=p_n\frac{p_{\nu_1}\dots p_{\nu_N}}{p_{\mu_1}\dots p_{\mu_N}}|\mel{n}{M_{\nu_1\mu_1}\dots M_{\nu_N\mu_N}}{m}|^2\notag\\
	&=p_np_{\nu_1}\dots p_{\nu_N}|\mel{n}{\mel{\nu_1}{U}{\mu_1}\dots\mel{\nu_N}{U}{\mu_N}}{m}|^2.\label{eq:path.prob.2}
\end{align}
By inserting the relation $\sum_{k_i}\dyad{k_i}=\mds{1}_S$, we can expand these terms as follows:
\begin{align}
	|\mel{m}{\mel{\mu_N}{U}{\nu_N}\dots\mel{\mu_1}{U}{\nu_1}}{n}|^2&=\qty|\sum_{k_1,\dots,k_{N-1}}\mel{m,\mu_N}{U}{k_{N-1},\nu_N}\dots\mel{k_2,\mu_2}{U}{k_1,\nu_2}\mel{k_1,\mu_1}{U}{n,\nu_1}|^2\label{eq:expand.1},\\
	|\mel{n}{\mel{\nu_1}{U}{\mu_1}\dots\mel{\nu_N}{U}{\mu_N}}{m}|^2&=\qty|\sum_{k_1,\dots,k_{N-1}}\mel{n,\nu_1}{U}{k_1,\mu_1}\mel{k_1,\nu_2}{U}{k_2,\mu_2}\dots\mel{k_{N-1},\nu_N}{U}{m,\mu_N}|^2.\label{eq:expand.2}
\end{align}
Because the Hamiltonians $H_S$ and $H_E$ are nondegenerate, there is at most one sequence $(k_1,\dots,k_{N-1})$ for which the terms inside the summation in Eqs.~\eqref{eq:expand.1} and \eqref{eq:expand.2} are nonzero.
Therefore, using the symmetry $|\mel{m,\mu}{U}{n,\nu}|=|\mel{n,\nu}{U}{m,\mu}|$, we can show that
\begin{align}
	|\mel{n}{\mel{\nu_1}{U}{\mu_1}\dots\mel{\nu_N}{U}{\mu_N}}{m}|^2&=|\mel{n,\nu_1}{U}{k_1,\mu_1}\mel{k_1,\nu_2}{U}{k_2,\mu_2}\dots\mel{k_{N-1},\nu_N}{U}{m,\mu_N}|^2\notag\\
	&=|\mel{k_1,\mu_1}{U}{n,\nu_1}\mel{k_2,\mu_2}{U}{k_1,\nu_2}\dots\mel{m,\mu_N}{U}{k_{N-1},\nu_N}|^2\notag\\
	&=|\mel{m,\mu_N}{U}{k_{N-1},\nu_N}\dots\mel{k_2,\mu_2}{U}{k_1,\nu_2}\mel{k_1,\mu_1}{U}{n,\nu_1}|^2\notag\\
	&=|\mel{m}{\mel{\mu_N}{U}{\nu_N}\dots\mel{\mu_1}{U}{\nu_1}}{n}|^2.
\end{align}
Combining this equality with Eqs.~\eqref{eq:path.prob.1} and \eqref{eq:path.prob.2} immediately yields $\mds{P}(\gamma)=\widetilde{\mds{P}}(\gamma)$.

We note that the assumptions of $|\mca{S}_{m\mu}|\le 2$ and nondegeneracy of Hamiltonians are crucial for the derivation above. To clarify their necessity, we demonstrate below that violating these assumptions can indeed lead to a positive asymmetry.
Due to energy conservation, the total unitary operator $U$ adopts a block-diagonal form:
\begin{equation}
	U=\bigoplus_\epsilon U_\epsilon,
\end{equation}
where $\epsilon$ denotes the conserved total energy of the system and environment within each block and $U_\epsilon$ is the corresponding unitary operator.
Under the stated assumptions, each block $U_\epsilon$ is guaranteed to have a maximum dimension of two, which ensures the forward-backward symmetry of the transition probabilities. 
This fact can be easily derived from the unitarity conditions $\sum_{i}|\mel{i}{U}{j}|^2=\sum_{j}|\mel{i}{U}{j}|^2=1$ as $|\mel{1}{U}{2}|^2=|\mel{2}{U}{1}|^2~(=1-|\mel{1}{U}{1}|^2)$.
However, if these assumptions are not satisfied, it is possible for some blocks to have dimension greater than two. In such cases, symmetry can be broken; in particular, one may find $|\mel{m,\mu}{U}{n,\nu}|^2 \neq |\mel{n,\nu}{U}{m,\mu}|^2 $ for some eigenstates $\ket{m,\mu}$ and $\ket{n,\nu}$ belonging to the same energy block.
Consequently, $\mds{P}(\gamma)=\widetilde{\mds{P}}(\gamma)$ becomes invalid.
As a concrete counterexample, consider the following $3 \times 3$ unitary matrix $U_\epsilon$ expressed in the energy eigenbasis:
\begin{equation}
	U_\epsilon =\begin{pmatrix}
			\dfrac{4}{9} & -\dfrac{4}{9} & -\dfrac{7}{9}\\[2ex]
			\dfrac{8}{9} & \dfrac{1}{9} & \dfrac{4}{9}\\[2ex]
			-\dfrac{1}{9} & -\dfrac{8}{9} & \dfrac{4}{9}
		\end{pmatrix}.
\end{equation}
It is straightforward to verify that this matrix satisfies unitarity, yet generally violates the symmetry condition on transition probabilities between states (i.e., $|\mel{i}{U_\epsilon}{j}|^2\neq|\mel{j}{U_\epsilon}{i}|^2$ for $i\neq j$). This demonstrates that the asymmetry $\Sigma_*$ can become positive when the assumptions are relaxed.

The symmetry breaking above can also be understood intuitively from the perspective of the path structure generated by the underlying Hamiltonian.
In the two-dimensional case (i.e., $|\mca{S}_{m\mu}|\le 2$), there is only a single {\it skeleton} path connecting $\ket{1}$ to $\ket{2}$, namely $1\to 2$, once all loops (i.e., $1\to 1$, $2\to 2$, and $1\to 2\to 1$) are recursively excluded from the path. This uniqueness of the skeleton path implies that $\mel{1}{U}{2}=a\mel{1}{H}{2}$ and $\mel{2}{U}{1}=a\mel{2}{H}{1}$ for some weight $a$, thereby preserving the symmetry.
In contrast, for higher-dimensional cases (i.e., $|\mca{S}_{m\mu}|>2$), multiple skeleton paths connect $\ket{1}$ to $\ket{2}$, such as $1\to 2$ and $1\to 3\to 2$.
These distinct paths contribute with different weights and interfere with one another, which generically breaks the symmetry.
For example, in the three-dimensional case, $\mel{1}{U}{2}=a\mel{1}{H}{2}+b\mel{1}{H}{3}\mel{3}{H}{2}$ and $\mel{2}{U}{1}=a\mel{2}{H}{1}+b\mel{2}{H}{3}\mel{3}{H}{1}$ for some weights $a$ and $b$, so that the symmetry $|\mel{1}{U}{2}|=|\mel{2}{U}{1}|$ is no longer guaranteed due to the interference between the paths $1\leftrightarrow 2$ and $1\leftrightarrow 3\leftrightarrow 2$.
Furthermore, when the Hamiltonians $H_S$ and $H_E$ are degenerate, multiple trajectories $(k_1,\dots,k_{N-1})$ may remain on the right-hand sides of Eqs.~\eqref{eq:expand.1} and \eqref{eq:expand.2}.
Interference between these trajectories again occurs---an effect that can be interpreted as ``enhanced fluctuations'' among different trajectories---and ultimately leads to the asymmetry.

\subsection{Vanishing of $\Sigma_*$ for separability-preserving dynamics}
Here, we show that the asymmetry vanishes (i.e., $\Sigma_* = 0$) under separability-preserving dynamics, namely those that cannot generate quantum entanglement from any initial pure product state. In this sense, the ability to generate quantum entanglement is a necessary condition for yielding a nonzero asymmetry. 
However, it is worth noting that it is not a sufficient condition.

In general, using the operator Schmidt decomposition, the interaction Hamiltonian $H_I$ can always be expressed in the following form:
\begin{equation}\label{eq:gen.form.Hint}
	H_I=\sum_k\lambda_kV_{S,k}\otimes V_{E,k},
\end{equation}
where $\lambda_k\ge 0$, $V_{S,k}$ and $V_{E,k}$ are traceless Hermitian operators (as any nonzero trace contributions can be absorbed into the local terms $H_S$ and $H_E$) and satisfy the normalization condition $\tr(V_{S,k}^\dagger V_{S,k})=\tr(V_{E,k}^\dagger V_{E,k})=\delta_{kk'}$.
Now suppose the dynamics is separability-preserving, meaning it maps all pure product states to pure product states at all times. Then, for any initial pure product state $\ket{\psi_S}\otimes\ket{\phi_E}$, the time-evolved state remains factorized:
\begin{equation}
	U_t\ket{\psi_S,\phi_E}=\ket{\psi_S(t),\phi_E(t)},
\end{equation}
where $\ket{\psi,\phi}\equiv\ket{\psi}\otimes\ket{\phi}$ and $U_t\coloneqq e^{-iHt}$ is the full unitary evolution under the total Hamiltonian $H=H_S\otimes\mds{1}_E + H_I + \mds{1}_S\otimes H_E$.
In what follows, we will show that this constraint implies $H_I=0$, namely, there is no interaction between the system and the environment.
To this end, we take the time derivative of the equation above and obtain
\begin{equation}\label{eq:prod.state.evol}
	\frac{d}{dt}\ket{\psi_S(t),\phi_E(t)}=-iH\ket{\psi_S(t),\phi_E(t)}.
\end{equation}
Let $\ket{\psi_S(t)^\perp}$ and $\ket{\phi_E(t)^\perp}$ be arbitrary pure states orthogonal to $\ket{\psi_S(t)}$ and $\ket{\phi_E(t)}$, respectively.
Multiplying Eq.~\eqref{eq:prod.state.evol} from the left by $\bra{\psi_S(t)^\perp,\phi_E(t)^\perp}$, we can calculate as follows:
\begin{align}
	-i\mel{\psi_S(t)^\perp,\phi_E(t)^\perp}{H}{\psi_S(t),\phi_E(t)}&=\bra{\psi_S(t)^\perp,\phi_E(t)^\perp}\frac{d}{dt}\ket{\psi_S(t),\phi_E(t)}\notag\\
	&=\bra{\psi_S(t)^\perp}\frac{d}{dt}\ket{\psi_S(t)}\braket{\phi_E(t)^\perp}{\phi_E(t)}+\braket{\psi_S(t)^\perp}{\psi_S(t)}\bra{\phi_E(t)^\perp}\frac{d}{dt}\ket{\phi_E(t)}\notag\\
	&=0.
\end{align}
Thus, $\mel{\psi_S^\perp,\phi_E^\perp}{H}{\psi_S,\phi_E}=0$ for all choices of orthogonal pure states $\ket{\psi_S}$, $\ket{\phi_E}$, $\ket{\psi_S^\perp}$, and $\ket{\phi_E^\perp}$. Since $H=H_S\otimes\mds{1}_E + H_I + \mds{1}_S\otimes H_E$ and the local terms do not contribute to this matrix element, we readily obtain
\begin{equation}\label{eq:nec.cond}
	\mel{\psi_S^\perp,\phi_E^\perp}{H_I}{\psi_S,\phi_E}=0.
\end{equation}
Next, we express the interaction Hamiltonian in the following operator form:
\begin{equation}\label{eq:Hint.exp}
	H_I=\sum_{m,n}\dyad{m}{n}\otimes O_{mn},
\end{equation}
where the operators $\{O_{mn}\}$ acting on the environment are given by
\begin{align}
	O_{mn}=(\bra{m}\otimes\mds{1}_E)H_I(\ket{n}\otimes\mds{1}_E)=\sum_k\lambda_k\mel{m}{V_{S,k}}{n}V_{E,k}.
\end{align}
Since each $V_{E,k}$ is traceless, it follows that $\tr O_{mn}=0$ for all $m$ and $n$.
We now exploit Eq.~\eqref{eq:nec.cond} to derive further constraints on $H_I$.
For $m\neq n$, substituting $\ket{\psi_S}=\ket{n}$ and $\bra{\psi_S^\perp}=\bra{m}$ to Eq.~\eqref{eq:nec.cond} yields
\begin{equation}
	\mel{\phi_E^\perp}{O_{mn}}{\phi_E}=0.
\end{equation}
Since this equality holds for arbitrary orthogonal states $\ket{\phi_E}$ and $\ket{\phi_E^\perp}$, the operator $O_{mn}$ must be proportional to the identity, i.e., $O_{mn}=c_{mn}\mds{1}_E~\forall m\neq n$ for some constant $c_{mn}$.
However, as $\tr O_{mn}=0$, we conclude $c_{mn}=0$, and hence $O_{mn}=0$ for all $m\neq n$.
Again, by substituting $\ket{\psi_S}=(\ket{m}+\ket{n})/\sqrt{2}$ and $\ket{\psi_S^\perp}=(\ket{m}-\ket{n})/\sqrt{2}$ to Eq.~\eqref{eq:nec.cond}, we obtain
\begin{equation}
	\frac{1}{2}\mel{\phi_E^\perp}{O_{mm}-O_{nn}}{\phi_E}=0,
\end{equation}
which holds for arbitrary orthogonal states $\ket{\phi_E}$ and $\ket{\phi_E^\perp}$.
As a result, $O_{nn}=O_{11}+z_n\mds{1}_E$ for some constant $z_n$.
The tracelessness $\tr O_{nn}=0$ implies $z_n=0$, so $O_{nn}=O_{11}$ for all $n$.
Inserting these expressions of $O_{mn}$ into Eq.~\eqref{eq:Hint.exp}, we obtain
\begin{align}
	H_I=\sum_n\dyad{n}\otimes O_{11} = \mds{1}_S\otimes O_{11}.
\end{align}
By expressing the interaction Hamiltonian as $H_I=\sum_{\mu,\nu}Q_{\mu\nu}\otimes\dyad{\mu}{\nu}$, where $\{Q_{\mu\nu}\}$ are traceless operators acting on the system, and repeating the above procedure in the environmental basis, we similarly obtain
\begin{equation}
	H_I=Q_{11}\otimes\mds{1}_E.
\end{equation}
Combining this with the earlier expression $H_I=\mds{1}_S\otimes O_{11}$, we conclude that $\mds{1}_S\otimes O_{11}=Q_{11}\otimes\mds{1}_E$.
Since both $O_{11}$ and $Q_{11}$ are traceless and $\tr_S(A_S\otimes B_E)=\tr_S(A_S)B_E$, this identity implies
\begin{align}
	O_{11}&=\frac{1}{\tr\mds{1}_S}\tr_S(\mds{1}_S\otimes O_{11})=\tr_S(Q_{11}\otimes\mds{1}_E)=0,\\
	Q_{11}&=\frac{1}{\tr\mds{1}_E}\tr_E(Q_{11}\otimes\mds{1}_E)=\tr_E(\mds{1}_S\otimes O_{11})=0.
\end{align}
Hence, we arrive at $H_I = 0$, demonstrating that separability-preserving dynamics preclude any genuine system-environment interaction. 
For simplicity, we focus on scenario (a), where the system and environment are in contact over a fixed time interval $T$. The same conclusion holds straightforwardly for scenario (b).
Since the interaction Hamiltonian vanishes in this case, any nondegenerate stationary state $\varrho_S$ of the system must satisfy $[\varrho_S,H_S]=0$.
Using this relation, the path probabilities can be calculated as follows:
\begin{align}
	\mds{P}(\gamma)&=p_np_\nu|\mel{m,\mu}{U}{n,\nu}|^2=p_np_\nu|\mel{m}{e^{-iH_ST}}{n}|^2|\mel{\mu}{e^{-iH_ET}}{\nu}|^2=p_np_\nu \delta_{mn}\delta_{\mu\nu},\\
	\widetilde{\mds{P}}(\gamma)&=p_np_\nu|\mel{n,\nu}{U}{m,\mu}|^2=p_np_\nu|\mel{n}{e^{-iH_ST}}{m}|^2|\mel{\nu}{e^{-iH_ET}}{\mu}|^2=p_np_\nu \delta_{mn}\delta_{\mu\nu}.
\end{align}
This verifies that $\mds{P}(\gamma)=\widetilde{\mds{P}}(\gamma)$, and so $\Sigma_*=0$.

\subsection{Vanishing of $\Sigma_*$ in incoherent Markovian dynamics}
We demonstrate that $\Sigma_*$ vanishes for stationary Markovian dynamics in the absence of quantum coherence.
In the weak coupling regime, the time evolution of the system's density matrix is governed by the GKSL equation \cite{Lindblad.1976.CMP,Gorini.1976.JMP},
\begin{align}
	\dot\varrho_t&=\mca{L}(\varrho_t),\notag\\
	\mca{L}(\circ)&\coloneqq -i[H,\circ]+\sum_{k\ge 1}\qty(L_k\circ L_k^\dagger - \{L_k^\dagger L_k,\circ\}/2).
\end{align}
Here, $H$ and $\{L_k\}_{k\ge 1}$ are the Hamiltonian and jump operators, respectively.
We assume that the jump operators satisfy the local detailed balance condition.
That is, the jump operators come in pairs $(k,k^*)$ such that $L_k=e^{\Delta s_k/2}L_{k^*}^\dagger$, where $\Delta s_k$ denotes the environmental entropy change due to the $k$th jump.
It is allowed that $k^*=k$, which implies that the jump operator $L_k$ is Hermitian.
For the short time interval $dt\ll 1$, the GKSL equation can be expressed in the form of the Kraus representation as 
\begin{equation}
	\varrho_{t+dt}=\sum_{k\ge 0}\mca{J}_k^\dagger\varrho_t\mca{J}_k,
\end{equation}
where $\mca{J}_0\coloneqq\mds{1}_S-iH_{\rm eff}dt$ and $\mca{J}_k\coloneqq L_k\sqrt{dt}$ for $k\ge 1$.
The operator $\mca{J}_0$, governed by the effective non-Hermitian Hamiltonian $H_{\rm eff}\coloneqq H-(i/2)\sum_{k\ge 1}L_k^\dagger L_k$, represents the case where no jump occurs.
On the other hand, the operator $\mca{J}_k$ characterizes the dynamics when the $k$th jump is detected.
For each trajectory $\gamma=\{n,(t_1,k_1),\dots,(t_N,k_N),m\}$, where the $k_i$th jump occurs at time $t_i$ for $1\le i\le N$, the probability of observing the trajectory $\gamma$ is given by
\begin{align}
	\mds{P}(\gamma)&=p_n|\mel{m}{U_{\rm eff}({T}-t_N)\mca{J}_{k_N}\dots\mca{J}_{k_1}U_{\rm eff}(t_1)}{n}|^2,
\end{align}
where $U_{\rm eff}(t)\coloneqq e^{-iH_{\rm eff}t}=e^{(-iH-\sum_{k\ge 1}L_k^\dagger L_k/2)t}$.

We define the backward process, where the Hamiltonian and jump operators are the time-reversed counterparts in the forward dynamics,
\begin{equation}
	\widetilde{H}=\Theta_SH\Theta_S^\dagger,~\widetilde{L}_k=\Theta_SL_k\Theta_S^\dagger.
\end{equation}
The effective Hamiltonian becomes $\widetilde{H}_{\rm eff}=\widetilde{H}-(i/2)\sum_{k\ge 1}\widetilde{L}_k^\dagger\widetilde{L}_k=\Theta_S[H+(i/2)\sum_{k\ge 1}L_k^\dagger L_k]\Theta_S^\dagger$.
The probability of observing the time-reversed trajectory $\widetilde{\gamma}=\{m,({T}-t_N,k_N^*),\dots,({T}-t_1,k_1^*),n\}$ in the backward process is given by
\begin{align}
	\widetilde{\mds{P}}(\widetilde{\gamma})&=p_m|\mel{n}{\Theta_S^\dagger\widetilde{U}_{\rm eff}(t_1)\widetilde{\mca{J}}_{k_1^*}\dots \widetilde{\mca{J}}_{k_N^*}\widetilde{U}_{\rm eff}({T}-t_N)\Theta_S}{m}|^2,
\end{align}
where $\widetilde{U}_{\rm eff}(t)\coloneqq e^{-i\widetilde{H}_{\rm eff}t}=\Theta_Se^{(iH-\sum_{k\ge 1}L_k^\dagger L_k/2)t}\Theta_S^\dagger$ and $\widetilde{\mca{J}}_k\coloneqq\widetilde{L}_k\sqrt{dt}$.
It can be confirmed that the entropy production $\Sigma$ can be expressed in terms of quantum states and path probabilities as
\begin{equation}
	\Sigma=\tr(\varrho_0\ln\varrho_0)-\tr(\varrho_{{T}}\ln\varrho_{{T}})+\int_0^\tau\dd{t}\sum_{k\ge 1}\tr(L_k\varrho_tL_k^\dagger)\Delta s_k=\ev{\ln\frac{\mds{P}(\gamma)}{\widetilde{\mds{P}}(\widetilde{\gamma})}}.
\end{equation}

Now, we are ready to show that $\mds{P}(\gamma)=\widetilde{\mds{P}}(\gamma)$ for the incoherent case, where the Hamiltonian and the jump operators are given by $H=\sum_n\epsilon_n\dyad{n}$ and $L_k=\sqrt{w_{mn}}\dyad{m}{n}$.
For convenience, we define $\msf{L}\coloneqq \sum_{k\ge 1}L_k^\dagger L_k$.
Noting that $[H,\msf{L}]=0$, we can calculate the forward probability as follows:
\begin{align}
	\mds{P}(\gamma)&=p_n|\mel{m}{U_{\rm eff}({T}-t_N)\mca{J}_{k_N}\dots\mca{J}_{k_1}U_{\rm eff}(t_1)}{n}|^2\notag\\
	&=p_n|\mel{m}{e^{-iH({T}-t_N)}e^{-\msf{L}({T}-t_N)/2}\mca{J}_{k_N}e^{-iH(t_N-t_{N-1})}e^{-\msf{L}(t_N-t_{N-1})/2}\dots e^{-iH(t_2-t_1)}e^{-\msf{L}(t_2-t_1)/2}\mca{J}_{k_1}e^{-iHt_1}e^{-\msf{L}t_1/2}}{n}|^2\notag\\
	&=p_n|\mel{m}{e^{-\msf{L}({T}-t_N)/2}\mca{J}_{k_N}e^{-\msf{L}(t_N-t_{N-1})/2}\dots e^{-\msf{L}(t_2-t_1)/2}\mca{J}_{k_1}e^{-\msf{L}t_1/2}}{n}|^2.
\end{align}
Here, we use the facts that $\bra{n}e^{-iHt}=\bra{n}e^{-i\epsilon_nt}$, $\mca{J}_ke^{-iHt}=e^{-i\epsilon_nt}\mca{J}_k$ for $L_k=\sqrt{w_{mn}}\dyad{m}{n}$, and the phase factors do not contribute to the probability.
Similarly, we can also show that
\begin{align}
	\widetilde{\mds{P}}(\gamma)&=p_n|\mel{m}{\Theta_S^\dagger\widetilde{U}_{\rm eff}({T}-t_N)\widetilde{\mca{J}}_{k_N}\dots \widetilde{\mca{J}}_{k_1}\widetilde{U}_{\rm eff}(t_1)\Theta_S}{n}|^2\notag\\
	&=p_n|\mel{m}{e^{-\msf{L}({T}-t_N)/2}\mca{J}_{k_N}e^{-\msf{L}(t_N-t_{N-1})/2}\dots e^{-\msf{L}(t_2-t_1)/2}\mca{J}_{k_1}e^{-\msf{L}t_1/2}}{n}|^2.
\end{align}
Therefore, $\mds{P}(\gamma)=\widetilde{\mds{P}}(\gamma)$, and the forward-backward asymmetry $\Sigma_*$ vanishes.

\section{Generalized quantum uncertainty relations for Markovian dynamics}
Here, we demonstrate the application of our results to Markovian dynamics.

\subsection{Generalized quantum TUR}
First, we apply the result (\FirRes) to derive a generalized quantum TUR for Markovian dynamics that satisfy the local detailed balance condition.
The generalized quantum TUR reads
\begin{equation}
	\frac{\mvar[\phi]}{\ev{\phi}^2}\ge f(\Sigma+\Sigma_*),
\end{equation}
where the forward-backward asymmetry for Markovian cases can be expressed as follows:
\begin{align}
	\Sigma_*&=\ev{\ln\frac{p_n|\mel{m}{U_{\rm eff}({T}-t_N)\mca{J}_{k_N}\dots\mca{J}_{k_1}U_{\rm eff}(t_1)}{n}|^2}{p_n|\mel{m}{\Theta_S^\dagger\widetilde{U}_{\rm eff}({T}-t_N)\widetilde{\mca{J}}_{k_N}\dots \widetilde{\mca{J}}_{k_1}\widetilde{U}_{\rm eff}(t_1)\Theta_S}{n}|^2}}\notag\\
	&=\ev{\ln\frac{|\mel{m}{U_{\rm eff}({T}-t_N)\mca{J}_{k_N}\dots\mca{J}_{k_1}U_{\rm eff}(t_1)}{n}|^2}{|\mel{m}{U_{\rm eff}({T}-t_N)^\dagger\mca{J}_{k_N}\dots \mca{J}_{k_1}U_{\rm eff}(t_1)^\dagger}{n}|^2}}.
\end{align}
Here, we use the relation $\Theta_S^\dagger\widetilde{U}_{\rm eff}(t)\Theta_S=e^{(iH-\sum_{k\ge 1}L_k^\dagger L_k/2)t}=U_{\rm eff}(t)^\dagger$ to obtain the last line.
As can be seen, the asymmetry between the forward and backward processes originates from replacing the Hamiltonian $H$ with $-H$.

\subsubsection{Short-time behavior of $\Sigma_*$}
We investigate the asymptotic behavior of $\Sigma_*$ in the short-time limit ${T}\ll 1$. In this regime, the asymmetry is dominated by paths with at most one jump.
Therefore, $\Sigma_*$ can be approximated as follows:
\begin{align}
	\Sigma_*&=\sum_{m,n}p_n|\mel{m}{U_{\rm eff}({T})}{n}|^2\ln\frac{|\mel{m}{U_{\rm eff}({T})}{n}|^2}{|\mel{m}{U_{\rm eff}({T})^\dagger}{n}|^2}\notag\\
	&+\sum_{m,n,k}\int_0^{{T}}\dd{t}p_n|\mel{m}{U_{\rm eff}({T}-t)L_kU_{\rm eff}(t)}{n}|^2\ln\frac{|\mel{m}{U_{\rm eff}({T}-t)L_kU_{\rm eff}(t)}{n}|^2}{|\mel{m}{U_{\rm eff}({T}-t)^\dagger L_kU_{\rm eff}(t)^\dagger}{n}|^2}+O({T}^3).
\end{align}
To evaluate the first term that is contributed by no-jump paths, we apply the following approximations:
\begin{align}
	|\mel{m}{U_{\rm eff}({T})}{n}|^2&=|\delta_{mn}-\mel{m}{iH_{\rm eff}}{n}{T}-\mel{m}{H_{\rm eff}^2}{n}{T}^2/2|^2+O({T}^3)\notag\\
	&=\delta_{mn}\qty[1-\mel{n}{iH_{\rm eff}-iH_{\rm eff}^\dagger}{n}{T}-\mel{n}{H_{\rm eff}^2+(H_{\rm eff}^\dagger)^2}{n}{T}^2]+|\mel{m}{iH_{\rm eff}}{n}|^2{T}^2+O({T}^3),\\
	|\mel{m}{U_{\rm eff}({T})^\dagger}{n}|^2&=\delta_{mn}\qty[1-\mel{n}{iH_{\rm eff}-iH_{\rm eff}^\dagger}{n}{T}-\mel{n}{H_{\rm eff}^2+(H_{\rm eff}^\dagger)^2}{n}{T}^2]+|\mel{m}{iH_{\rm eff}^\dagger}{n}|^2{T}^2+O({T}^3),
\end{align}
where we use the expansion $U_{\rm eff}(t)=\mds{1}_S-iH_{\rm eff}t-H_{\rm eff}^2t^2/2+O(t^3)$ for $t\ll 1$.
These yield the following approximation for the first term:
\begin{equation}\label{eq:first.approx}
	(\text{1st})={T}^2\sum_{m,n}p_n|\mel{m}{iH_{\rm eff}}{n}|^2\ln\frac{|\mel{m}{iH_{\rm eff}}{n}|^2}{|\mel{m}{iH_{\rm eff}^\dagger}{n}|^2}+O({T}^3).
\end{equation}
To evaluate the second term contributed by one-jump paths, we use the following approximations:
\begin{align}
	&|\mel{m}{U_{\rm eff}({T}-t)L_kU_{\rm eff}(t)}{n}|^2\notag\\
	&=|\mel{m}{(\mds{1}_S-iH_{\rm eff}({T}-t))L_k(\mds{1}_S-iH_{\rm eff}t)}{n}|^2+O({T}^2)\notag\\
	&=\mel{m}{(\mds{1}_S-iH_{\rm eff}({T}-t))L_k(\mds{1}_S-iH_{\rm eff}t)}{n}\mel{n}{(\mds{1}_S+iH_{\rm eff}^\dagger t)L_k^\dagger(\mds{1}_S+iH_{\rm eff}^\dagger({T}-t))}{m}+O({T}^2)\notag\\
	&=|\mel{m}{L_k}{n}|^2+2\Re[\mel{m}{L_k}{n}\mel{n}{iH_{\rm eff}^\dagger L_k^\dagger}{m}]t+2\Re[\mel{m}{L_k}{n}\mel{n}{iL_k^\dagger H_{\rm eff}^\dagger}{m}]({T}-t)+O({T}^2),\\
	&|\mel{m}{U_{\rm eff}({T}-t)^\dagger L_kU_{\rm eff}(t)^\dagger}{n}|^2\notag\\
	&=|\mel{m}{L_k}{n}|^2+2\Re[\mel{m}{L_k}{n}\mel{n}{-iH_{\rm eff} L_k^\dagger}{m}]t+2\Re[\mel{m}{L_k}{n}\mel{n}{-iL_k^\dagger H_{\rm eff}}{m}]({T}-t)+O({T}^2).
\end{align}
Applying $\ln(1+x)=x+O(x^2)$ for $x\ll 1$, the second term can be approximated as
\begin{align}
	(\text{2nd})&=2\sum_{m,n,k}\int_0^{{T}}\dd{t}p_n\Big\{\Re[\mel{m}{L_k}{n}\mel{n}{iH_{\rm eff}^\dagger L_k^\dagger}{m}+\mel{m}{L_k}{n}\mel{n}{iH_{\rm eff} L_k^\dagger}{m}]t\notag\\
	&\hspace{2.75cm}+\Re[\mel{m}{L_k}{n}\mel{n}{iL_k^\dagger H_{\rm eff}^\dagger}{m}+\mel{m}{L_k}{n}\mel{n}{iL_k^\dagger H_{\rm eff}}{m}]({T}-t)\Big\}+O({T}^3)\notag\\
	&={T}^2\sum_{k}\qty{\Re[i\tr(L_k\varrho_SH_{\rm eff}^\dagger L_k^\dagger+L_k\varrho_SH_{\rm eff}L_k^\dagger)]+\Re[i\tr(L_k\varrho_SL_k^\dagger H_{\rm eff}^\dagger+L_k\varrho_SL_k^\dagger H_{\rm eff})]}+O({T}^3)\notag\\
	&=2{T}^2\sum_{k}\qty{\Re[i\tr(L_k\varrho_S H L_k^\dagger)]+\Re[i\tr(L_k\varrho_SL_k^\dagger H)]}+O({T}^3)\notag\\
	&={T}^2i\tr(\varrho_S[H,\msf{L}])+O({T}^3).\label{eq:sec.approx}
\end{align}
Here we use the facts that $2\Re(z)=z+z^*$ and $\tr(L_k\varrho_SL_k^\dagger H)$ is a real number to obtain the last line.
By combining the approximations of these two terms [Eqs.~\eqref{eq:first.approx} and \eqref{eq:sec.approx}], we get the following expression for the forward-backward asymmetry:
\begin{align}
	\Sigma_*&={T}^2\qty[\sum_{m,n}p_n|\mel{m}{iH_{\rm eff}}{n}|^2\ln\frac{|\mel{m}{iH_{\rm eff}}{n}|^2}{|\mel{m}{iH_{\rm eff}^\dagger}{n}|^2}+i\tr(\varrho_S[H,\msf{L}])]+O({T}^3).
\end{align}
For simplicity, we define $\ev{A}_{S}\coloneqq\tr(A\varrho_S)$ hereafter.
By simple algebraic calculations, we can verify that
\begin{align}
	\sum_{m,n}p_n|\mel{m}{iH_{\rm eff}}{n}|^2&=\ev{H_{\rm eff}^\dagger H_{\rm eff}}_S,\\
	\sum_{m,n}p_n|\mel{m}{iH_{\rm eff}^\dagger}{n}|^2&=\ev{H_{\rm eff}H_{\rm eff}^\dagger}_S,\\
	\ev{H_{\rm eff}H_{\rm eff}^\dagger}_S-\ev{H_{\rm eff}^\dagger H_{\rm eff}}_S&=i\ev{[H,\msf{L}]}_S,\\
	\ev{H_{\rm eff}^\dagger H_{\rm eff}}_S+\ev{H_{\rm eff}H_{\rm eff}^\dagger}_S&=\ev{2H^2+\msf{L}^2/2}_S,
\end{align}
where $\msf{L}=\sum_{k\ge 1}L_k^\dagger L_k$.
Exploiting the convexity of function $x\ln(x/y)$ over $(0,+\infty)\times(0,+\infty)$, we can lower bound the leading-order term of $\Sigma_*$ as follows:
\begin{align}
	&\sum_{m,n}p_n|\mel{m}{iH_{\rm eff}}{n}|^2\ln\frac{|\mel{m}{iH_{\rm eff}}{n}|^2}{|\mel{m}{iH_{\rm eff}^\dagger}{n}|^2}+i\ev{[H,\msf{L}]}_S\notag\\
	&=\sum_{m,n}p_n|\mel{m}{iH_{\rm eff}}{n}|^2\ln\frac{|\mel{m}{iH_{\rm eff}}{n}|^2}{|\mel{m}{iH_{\rm eff}^\dagger}{n}|^2} + \ev{H_{\rm eff}H_{\rm eff}^\dagger}_S-\ev{H_{\rm eff}^\dagger H_{\rm eff}}_S\notag\\
	&\ge \ev{H_{\rm eff}^\dagger H_{\rm eff}}_S\ln\frac{\ev{H_{\rm eff}^\dagger H_{\rm eff}}_S}{\ev{H_{\rm eff}H_{\rm eff}^\dagger}_S}-\ev{H_{\rm eff}^\dagger H_{\rm eff}}_S+\ev{H_{\rm eff}H_{\rm eff}^\dagger}_S\notag\\
	&\ge c_*\frac{\qty(\ev{H_{\rm eff}^\dagger H_{\rm eff}}_S-\ev{H_{\rm eff}H_{\rm eff}^\dagger}_S)^2}{\ev{H_{\rm eff}^\dagger H_{\rm eff}}_S+\ev{H_{\rm eff}H_{\rm eff}^\dagger}_S}\notag\\
	&=c_*\frac{|\ev{[H,\msf{L}]}_S|^2}{\ev{2H^2+\msf{L}^2/2}_S}.
\end{align}
Here, we apply Jensen's inequality to obtain the third line and the following inequality \cite{Shiraishi.2016.PRL} to obtain the fourth line:
\begin{equation}
	x\ln\frac{x}{y}-x+y\ge c_*\frac{(x-y)^2}{x+y},
\end{equation}
where $c_*=8/9$.
Consequently, $\Sigma_*$ is lower bounded in the short-time regime as
\begin{equation}
	\Sigma_*\ge c_*\frac{|\ev{[H,\msf{L}]}_S|^2}{\ev{2H^2+\msf{L}^2/2}_S}{T}^2.
\end{equation}

\subsubsection{General lower bound of $\Sigma_*$}
Here we derive a lower bound of $\Sigma_*$ for arbitrary finite times.
To this end, let us recall that the backward probability can be calculated as follows:
\begin{align}
	\widetilde{\mds{P}}(\gamma)&=|\mel{m}{U_{\rm eff}({T}-t_N)^\dagger\mca{J}_{k_N}\dots \mca{J}_{k_1}U_{\rm eff}(t_1)^\dagger}{n}|^2,
\end{align}
where $U_{\rm eff}(t)^\dagger=e^{(iH-\sum_{k\ge 1}L_k^\dagger L_k/2)t}$.
By exploiting the data-processing inequality for the relative entropy, we obtain a lower bound on $\Sigma_*$ in terms of the Kullback-Leibler divergence as follows:
\begin{equation}\label{eq:asym.lb}
	\Sigma_*=\ev{\ln\frac{\mds{P}(\gamma)}{\widetilde{\mds{P}}(\gamma)}}\ge D(\vb*{p}||\vb*{q}),
\end{equation}
where $\vb*{p}=[p_1,\dots,p_d]^\top$, $\vb*{q}=[q_1,\dots,q_d]^\top$, $q_n=\mel{n}{\widetilde{\varrho}_S({T})}{n}$, and $\widetilde{\varrho}_S({T})= e^{\widetilde{\mca{L}}{T}}(\varrho_{S})$ is the final state of a modified process governed by the following superoperator:
\begin{equation}
	\widetilde{\mca{L}}(\circ)\coloneqq i[H,\circ]+\sum_{k\ge 1}(L_k\circ L_k^\dagger - \{L_k^\dagger L_k,\circ \}/2).
\end{equation}
The lower bound in Eq.~\eqref{eq:asym.lb} characterizes the difference between the final state of the original dynamics and that of the modified dynamics in which the Hamiltonian $H$ is replaced by $-H$.

As can be observed, the original superoperator $\mca{L}$ and the modified one $\widetilde{\mca{L}}$ differ only in the unitary part. This indicates that the lower bound $D(\vb*{p}||\vb*{q})$ of the asymmetry originates from quantum coherence inherently present in the unitary evolution. In the absence of quantum coherence, the Hamiltonian commutes with the system's quantum state at all times, causing the unitary contribution to vanish entirely. In this case, the original and modified superoperator become identical, i.e., $\mca{L}\equiv\widetilde{\mca{L}}$. As a result, we find $\widetilde{\varrho}_S(T)=e^{\widetilde{\mca{L}}{T}}(\varrho_{S})=e^{\mca{L}{T}}(\varrho_{S})=\varrho_S$, leading to $\vb*{p}=\vb*{q}$ and hence $D(\vb*{p}||\vb*{q})=0$.

\subsection{Generalized quantum KUR}
Next, we establish a generalized quantum KUR for Markovian dynamics without assuming any specific conditions on the jump operators, such as the local detailed balance condition required in the derivation of the generalized quantum TUR. This formulation thus encompasses a broad class of physically relevant processes, including spontaneous emission and collective dissipation.
For such general Markovian dynamics, the generalized quantum KUR (\SecRes) reads
\begin{equation}
	\frac{\mvar[\phi]}{\ev{\phi}^2}\ge\frac{1}{\msc{P}^{-1}-1},
\end{equation}
where the inactivity term $\msc{P}$ is the probability of observing no jump in the system and can be calculated as
\begin{equation}
	\msc{P}=\tr(e^{-iH_{\rm eff}{T}}\varrho_Se^{iH_{\rm eff}^\dagger{T}}).
\end{equation}
For the short-time limit ${T}\ll 1$, $\msc{P}$ can be expressed as follows:
\begin{align}
	\msc{P}&=\tr{[\mds{1}_S-iH_{\rm eff}{T}+O({T}^2)]\varrho_S[\mds{1}_S+iH_{\rm eff}^\dagger{T}+O({T}^2)]}\notag\\
	&=1+\tr{(-iH_{\rm eff}+iH_{\rm eff}^\dagger)\varrho_S}{T}+O({T}^2)\notag\\
	&=1-{T}\sum_{k\ge 1}\tr(L_k\varrho_SL_k^\dagger)+O({T}^2)\notag\\
	&=1-\mca{A}_{{T}}+O({T}^2),
\end{align}
where $\mca{A}_{{T}}\coloneqq{T}\sum_{k\ge 1}\tr(L_k\varrho_SL_k^\dagger)$ is the dynamical activity over time ${T}$.
Therefore, the lower-bound term $\msc{P}^{-1}-1$ can be approximated as
\begin{align}
	\msc{P}^{-1}-1&=[1-\mca{A}_{{T}}+O({T}^2)]^{-1}-1\notag\\
	&=\mca{A}_{{T}}+O({T}^2),
\end{align}
which, to leading order, coincides with the dynamical activity.

In Ref.~\cite{Hasegawa.2021.PRL2}, the following uncertainty relation was derived for Markovian dynamics using the Loschmidt echo approach:
\begin{equation}
	\frac{\mvar[\phi]}{\ev{\phi}^2}\ge\frac{1}{\eta^{-1}-1}.
\end{equation}
Here, $\eta$ is explicitly given by
\begin{equation}
	\eta=\qty|\tr(e^{-iH_{\rm eff}{T}}\varrho_S)|^2.
\end{equation}
Applying the inequality $|\tr(AB)|^2\le\tr(A^\dagger A)\tr(B^\dagger B)$, we can show that $\eta$ is always smaller than $\msc{P}$ as follows:
\begin{equation}
	\eta=\qty|\tr(e^{-iH_{\rm eff}{T}}\sqrt{\varrho_S}\sqrt{\varrho_S})|^2\le\tr(e^{-iH_{\rm eff}{T}}\varrho_Se^{iH_{\rm eff}^\dagger{T}})\tr(\varrho_S)=\msc{P}.
\end{equation}
Therefore, our new relation (\SecRes) is tighter than the extant one,
\begin{equation}
	\frac{\mvar[\phi]}{\ev{\phi}^2}\ge\frac{1}{\msc{P}^{-1}-1}\ge\frac{1}{\eta^{-1}-1}.
\end{equation}

\section{Generalization of the main result (\FirRes) to arbitrary initial states}
Here we derive the generalization of the first main result (\FirRes) to arbitrary initial states.
To this end, we first briefly explain the setup for the general case.
Let $\varrho_S(0)=\sum_np_n\dyad{n}$ be the spectral decomposition of the initial state of the systems.
A projective measurement is initially performed on the system using the basis $\{\ket{n}\}$, which does not alter the system's state.
The system and the environment then evolve in the same way as in the stationary case, where the projective measurements are performed on the environment during each interaction.
Let $\varrho_S({T})=\sum_nq_n\dyad{n'}$ be the spectral decomposition of the final state of the system.
A projective measurement on the system is again performed at the final time using the basis $\{\ket{n'}\}$, which also do not alter the system's state.
For each stochastic trajectory $\gamma=\{n,(\nu_1,\mu_1),\dots,(\nu_N,\mu_N),m\}$ and its time-reversed (backward) counterpart $\widetilde{\gamma}=\{m,(\mu_N,\nu_N),\dots,(\mu_1,\nu_1),n\}$, the path probabilities of the forward and backward trajectories are given by
\begin{align}
	\mds{P}(\gamma)&=p_n|\mel{m'}{M_{\mu_N\nu_N}\dots M_{\mu_1\nu_1}}{n}|^2,\\
	\widetilde{\mds{P}}(\widetilde{\gamma})&=q_m|\mel{n}{\Theta_S^\dagger\widetilde{M}_{\nu_1\mu_1}\dots\widetilde{M}_{\nu_N\mu_N}\Theta_S}{m'}|^2.
\end{align}

Next, we show that entropy production can be expressed as the relative entropy between the forward and backward path probabilities.
Noting that
\begin{equation}
	\widetilde{\mds{P}}(\widetilde{\gamma})=\frac{q_m}{p_n}\frac{p_{\mu_1}\dots p_{\mu_N}}{p_{\nu_1}\dots p_{\nu_N}}\mds{P}(\gamma)=\frac{q_m}{p_n}e^{-\beta\sum_{i=1}^N(\epsilon_{\mu_i}-\epsilon_{\nu_i})}\mds{P}(\gamma),
\end{equation}
it can be easily shown that
\begin{align}
	\Sigma&=\Delta S+\beta\Delta Q\notag\\
	&=\ev{\ln p_n-\ln q_m+\beta\sum_{i=1}^N(\epsilon_{\mu_i}-\epsilon_{\nu_i})}\notag\\
	&=\sum_\gamma \mds{P}(\gamma)\ln\frac{\mds{P}(\gamma)}{\widetilde{\mds{P}}(\widetilde{\gamma})}\notag\\
	&=\ev{\ln\frac{\mds{P}(\gamma)}{\widetilde{\mds{P}}(\widetilde{\gamma})}}.
\end{align}
Additionally, the entropy production can also be expressed in terms of quantum states.
The average heat dissipation during each interaction can be similarly calculated as follows:
\begin{align}
	\sum_{\gamma}\mds{P}(\gamma)(\epsilon_{\mu_i}-\epsilon_{\nu_i})&=\sum_{m,n,\{\mu_j,\nu_j\}_{j=1}^N}p_n|\mel{m'}{M_{\mu_N\nu_N}\dots M_{\mu_1\nu_1}}{n}|^2(\epsilon_{\mu_i}-\epsilon_{\nu_i})\notag\\
	&=\sum_{\mu_i,\nu_i}(\epsilon_{\mu_i}-\epsilon_{\nu_i})\tr{M_{\mu_i\nu_i}\varrho_S[(i-1)\tau]M_{\mu_i\nu_i}^\dagger}\notag\\
	&=\sum_{\mu_i,\nu_i}p_{\nu_i}\epsilon_{\mu_i}\tr{\mel{\mu_i}{U}{\nu_i}\varrho_S[(i-1)\tau]\mel{\nu_i}{U^\dagger}{\mu_i}}-\sum_{\mu_i,\nu_i}p_{\nu_i}\epsilon_{\nu_i}\tr{\mel{\mu_i}{U}{\nu_i}\varrho_S[(i-1)\tau]\mel{\nu_i}{U^\dagger}{\mu_i}}\notag\\
	&=\tr{H_EU(\varrho_S[(i-1)\tau]\otimes\varrho_E)U^\dagger}-\tr{U(\varrho_S[(i-1)\tau]\otimes H_E\varrho_E)U^\dagger}\notag\\
	&=\tr{H_E\varrho_{SE}(i\tau)}-\tr{H_E\varrho_E},
\end{align}
where $\varrho_{SE}(i\tau)\coloneqq U(\varrho_S[(i-1)\tau]\otimes\varrho_E)U^\dagger$ and $\varrho_S(i\tau)\coloneqq\tr_E\varrho_{SE}(i\tau)$.
Following the same approach as in the stationary case, we obtain the following expression for the entropy production:
\begin{align}
	\Sigma&=\Delta S+\beta\Delta Q\notag\\
	&=\tr{\varrho_S(0)\ln\varrho_S(0)}-\tr{\varrho_S({T})\ln\varrho_S({T})}+\beta\sum_{i=1}^{N}\qty[\tr{H_E\varrho_{SE}(i\tau)}-\tr{H_E\varrho_E}]]\notag\\
	&=\sum_{i=1}^N\qty[\tr{\varrho_S[(i-1)\tau]\ln\varrho_S[(i-1)\tau]}-\tr{\varrho_S(i\tau)\ln\varrho_S(i\tau)}+\beta(\tr{H_E\varrho_{SE}(i\tau)}-\tr{H_E\varrho_E})]\notag\\
	&=\sum_{i=1}^N\qty[-\tr{\varrho_{SE}(i\tau)\ln\varrho_E}-\tr{\varrho_S(i\tau)\ln\varrho_S(i\tau)}+\tr{\varrho_S[(i-1)\tau]\ln\varrho_S[(i-1)\tau]}+\tr{\varrho_E\ln\varrho_E}]\notag\\
	&=\sum_{i=1}^N\qty[-\tr{\varrho_{SE}(i\tau)\ln[\varrho_S(i\tau)\otimes\varrho_E]}+\tr{(\varrho_S[(i-1)\tau]\otimes\varrho_E)\ln(\varrho_S[(i-1)\tau]\otimes\varrho_E)}]\notag\\
	&=\sum_{i=1}^N\qty[-\tr{\varrho_{SE}(i\tau)\ln[\varrho_S(i\tau)\otimes\varrho_E]}+\tr{\varrho_{SE}(i\tau)\ln\varrho_{SE}(i\tau)}]\notag\\
	&=\sum_{i=1}^N D(\varrho_{SE}(i\tau)\|\varrho_S(i\tau)\otimes\varrho_E).
\end{align}
Using the formulation of path probabilities
\begin{align}
	\mds{P}(\gamma)&=p_n|\mel{m'}{M_{\mu_N\nu_N}\dots M_{\mu_1\nu_1}}{n}|^2,\\
	\mds{P}(\widetilde{\gamma})&=p_m|\mel{n'}{M_{\nu_1\mu_1}\dots M_{\nu_N\mu_N}}{m}|^2,\\
	\widetilde{\mds{P}}(\widetilde{\gamma})&=q_m|\mel{n}{\Theta_S^\dagger\widetilde{M}_{\nu_1\mu_1}\dots\widetilde{M}_{\nu_N\mu_N}\Theta_S}{m'}|^2,\\
	\widetilde{\mds{P}}(\gamma)&=q_n|\mel{m}{\Theta_S^\dagger\widetilde{M}_{\mu_N\nu_N}\dots\widetilde{M}_{\mu_1\nu_1}\Theta_S}{n'}|^2,
\end{align}
we can show that
\begin{align}
	\widetilde{\mds{P}}(\widetilde{\gamma})&=q_m|\mel{n}{\Theta_S^\dagger\widetilde{M}_{\nu_1\mu_1}\dots\widetilde{M}_{\nu_N\mu_N}\Theta_S}{m'}|^2\notag\\
	&=q_m\frac{p_{\mu_1}\dots p_{\mu_N}}{p_{\nu_1}\dots p_{\nu_N}}|\mel{n}{M_{\mu_1\nu_1}^\dagger\dots M_{\mu_N\nu_N}^\dagger}{m'}|^2\notag\\
	&=q_m\frac{p_{\mu_1}\dots p_{\mu_N}}{p_{\nu_1}\dots p_{\nu_N}}|\mel{m'}{M_{\mu_N\nu_N}\dots M_{\mu_1\nu_1}}{n}|^2\notag\\
	&=\frac{q_m}{p_n}\frac{p_{\mu_1}\dots p_{\mu_N}}{p_{\nu_1}\dots p_{\nu_N}}\mds{P}(\gamma),\\
	\widetilde{\mds{P}}(\gamma)&=q_n|\mel{m}{\Theta_S^\dagger\widetilde{M}_{\mu_N\nu_N}\dots\widetilde{M}_{\mu_1\nu_1}\Theta_S}{n'}|^2\notag\\
	&=q_n\frac{p_{\nu_1}\dots p_{\nu_N}}{p_{\mu_1}\dots p_{\mu_N}}|\mel{m}{M_{\nu_N\mu_N}^\dagger\dots M_{\nu_1\mu_1}^\dagger}{n'}|^2\notag\\
	&=q_n\frac{p_{\nu_1}\dots p_{\nu_N}}{p_{\mu_1}\dots p_{\mu_N}}|\mel{n'}{M_{\nu_1\mu_1}\dots M_{\nu_N\mu_N}}{m}|^2\notag\\
	&=\frac{q_n}{p_m}\frac{p_{\nu_1}\dots p_{\nu_N}}{p_{\mu_1}\dots p_{\mu_N}}\mds{P}(\widetilde{\gamma}).
\end{align}
These relations immediately yield the following equality:
\begin{equation}
	\ln\frac{\widetilde{\mds{P}}(\widetilde{\gamma})}{\mds{P}(\widetilde{\gamma})}=\ln\frac{\mds{P}(\gamma)}{\widetilde{\mds{P}}(\gamma)}+\ln\frac{q_mq_n}{p_mp_n}.
\end{equation}
Consequently, the entropy production can be decomposed as follows:
\begin{align}
	\Sigma&=\ev{\ln\frac{\mds{P}(\gamma)}{\widetilde{\mds{P}}(\widetilde{\gamma})}}\notag\\
	&=\ev{\ln\frac{\mds{P}(\gamma)}{\mds{P}(\widetilde{\gamma})}-\ln\frac{\widetilde{\mds{P}}(\widetilde{\gamma})}{\mds{P}(\widetilde{\gamma})}}\notag\\
	&=\ev{\ln\frac{\mds{P}(\gamma)}{\mds{P}(\widetilde{\gamma})}-\ln\frac{\mds{P}(\gamma)}{\widetilde{\mds{P}}(\gamma)}-\ln\frac{q_mq_n}{p_mp_n}}\notag\\
	&=\ev{\ln\frac{\mds{P}(\gamma)}{\mds{P}(\widetilde{\gamma})}}-\Sigma_*-\mfr{b},
\end{align}
where we define the boundary term
\begin{equation}
	\mfr{b}\coloneqq\ev{\ln\frac{q_mq_n}{p_mp_n}}.
\end{equation}
Note that $\mfr{b}$ becomes negligible in the long-time regime compared with time-extensive quantities $\Sigma$ and $\Sigma_*$.
Following the same procedure as in the stationary case, we readily obtain the generalization of the main result (\FirRes) for arbitrary initial states:
\begin{equation}\label{eq:gen.tur}
	\frac{\mvar[\phi]}{\ev{\phi}^2}\ge f(\Sigma+\Sigma_*+\mfr{b}).
\end{equation}
In the stationary case, $\mfr{b}$ vanishes and the relation \eqref{eq:gen.tur} recovers the main result (\FirRes).

\section{Generalized TUR for underdamped Langevin dynamics}
Here we derive a generalized TUR for general underdamped Langevin dynamics and demonstrate the crucial role of the asymmetry in precision constraints.

\subsection{Setup and generalized TUR}
We consider an underdamped Langevin system consisting of $d$ particles, whose positions and velocities are denoted by $\vb*{x}=[x_1,\dots,x_d]^\top$ and $\vb*{v}=[v_1,\dots,v_d]^\top$, respectively.
Each particle $i\in\{1,\dots,d\}$ is coupled to an individual thermal reservoir at temperature $T_i$.
The system dynamics is governed by the following set of stochastic Langevin equations:
\begin{align}
	\dot{x}_i&=v_i,\\
	m_i\dot{v}_i&=F_i(\vb*{x},\vb*{v})-\gamma_i v_i+\xi_i,
\end{align}
where $m_i$ is the particle's mass, $\gamma_i$ the damping coefficient, $F_i$ the total force acting on the $i$th particle, and $\{\xi_i\}$ are Gaussian white noises satisfying $\ev{\xi_i}=0$ and $\ev{\xi_{i}(t)\xi_{j}(t')}=2\gamma_iT_i\delta_{ij}\delta(t-t')$.
Hereinafter, we set the Boltzmann constant to unity, $k_B=1$.
Note that the force terms $\{F_i(\vb*{x},\vb*{v})\}$ may include velocity-dependent contributions such as the Lorentz force. As a result, time-reversal symmetry can be broken, and the strong form of the detailed fluctuation theorem generally does not hold.

Let $p_t(\vb*{x},\vb*{v})$ denote the probability density of finding the system in state $(\vb*{x},\vb*{v})$ at time $t$.
Its evolution is governed by the Fokker-Planck equation:
\begin{equation}
	\dot p_t(\vb*{x},\vb*{v})= -\sum_{i=1}^d[\partial_{x_i}J_{x_i}(\vb*{x},\vb*{v},t)+\partial_{v_i}J_{v_i}(\vb*{x},\vb*{v},t)],
\end{equation}
where the probability currents are defined as
\begin{align}
	J_{x_i}(\vb*{x},\vb*{v},t)&=v_ip_t(\vb*{x},\vb*{v}),\\
	J_{v_i}(\vb*{x},\vb*{v},t)&=\frac{1}{m_i}\qty[-\gamma_iv_i+F_i(\vb*{x},\vb*{v})-\frac{\gamma_iT_i}{m_i}\partial_{v_i}]p_t(\vb*{x},\vb*{v}).
\end{align}
To discuss stochastic thermodynamics of this system, we decompose each force $F_i(\vb*{x},\vb*{v})$ into reversible and irreversible components: 
\begin{equation}
	F_i(\vb*{x},\vb*{v})=F_i^{\rm rev}(\vb*{x},\vb*{v})+F_i^{\rm irr}(\vb*{x},\vb*{v}),
\end{equation}
such that the relations $F_i^{\rm rev}(\vb*{x},\vb*{v})={F}_i^{\rm rev}(\vb*{x},-\vb*{v})^\dagger$ and $F_i^{\rm irr}(\vb*{x},\vb*{v})=-{F}_i^{\rm irr}(\vb*{x},-\vb*{v})^\dagger$ are fulfilled.
Here, $F^\dagger$ means that the signs of all odd parameters (e.g., magnetic fields) inside the function $F$ are reversed.
Consequently, the probability currents can also be split into reversible and irreversible parts as $J_{v_i}(\vb*{x},\vb*{v},t)=J_{v_i}^{\rm rev}(\vb*{x},\vb*{v},t)+J_{v_i}^{\rm irr}(\vb*{x},\vb*{v},t)$, where
\begin{align}
	J_{v_i}^{\rm rev}(\vb*{x},\vb*{v},t)&=\frac{1}{m_i}F_i^{\rm rev}(\vb*{x},\vb*{v})p_t(\vb*{x},\vb*{v}),\\
	J_{v_i}^{\rm irr}(\vb*{x},\vb*{v},t)&=\frac{1}{m_i}\qty[-\gamma_iv_i+F_i^{\rm irr}(\vb*{x},\vb*{v})-\frac{\gamma_iT_i}{m_i}\partial_{v_i}]p_t(\vb*{x},\vb*{v}).
\end{align}
Let $\Gamma = \{\vb*{x}(t),\vb*{v}(t)\}_{t=0}^{t=T}$ denote a stochastic trajectory of the system over the interval $[0,T]$. The probability of observing this trajectory in the forward process is given by
\begin{equation}
	\mds{P}(\Gamma)=p_0(\vb*{x}(0),\vb*{v}(0))\mds{P}(\Gamma|\vb*{x}(0),\vb*{v}(0)),
\end{equation}
where the conditional path probability takes the form
\begin{align}
	&\mds{P}(\Gamma|\vb*{x}(0),\vb*{v}(0))\notag\\
	&\propto\exp\qty(-\int_0^T\dd{t}\sum_{i=1}^d\qty{\frac{1}{4\gamma_iT_i}[m_i\dot v_i(t)-F_i(\vb*{x}(t),\vb*{v}(t))+\gamma_i v_i(t)]^2+\frac{1}{2m_i}\partial_{v_i}F_i(\vb*{x}(t),\vb*{v}(t))-\frac{\gamma_i}{2m_i}}).
\end{align}
Next, we construct the backward process, where the corresponding variables are denoted by $\widetilde{\vb*{x}}$ and $\widetilde{\vb*{v}}$.
The initial positions and velocities are sampled from the distribution $p_T(\widetilde{\vb*{x}},\widetilde{\vb*{v}})$, and their time evolution is subsequently governed by the following stochastic differential equation:
\begin{align}
	\dot{\widetilde{x}}_i&=\widetilde{v}_i,\\
	m_i\dot{\widetilde{v}}_i&={F}_i(\widetilde{\vb*{x}},\widetilde{\vb*{v}})^\dagger -\gamma_i \widetilde{v}_i+\xi_i.
\end{align}
As observed, the difference between the forward and backward processes stems from the time-reversal asymmetry of the forces $\{F_i\}$, especially in the presence of odd parameters (e.g., magnetic fields).
The probability of observing the time-reversed trajectory $\widetilde{\Gamma}=\{\vb*{x}(T-t),-\vb*{v}(T-t)\}_{t=0}^{t=T}$ in the backward process is given by $\widetilde{\mds{P}}(\widetilde{\Gamma})=p_T(\vb*{x}(T),\vb*{v}(T))\widetilde{\mds{P}}(\widetilde{\Gamma}|\vb*{x}(T),-\vb*{v}(T))$, where
\begin{align}
	&\widetilde{\mds{P}}(\widetilde{\Gamma}|\vb*{x}(T),-\vb*{v}(T))\notag\\
	&\propto\exp\qty(-\int_0^T\dd{t}\sum_{i=1}^d\qty{\frac{1}{4\gamma_iT_i}[m_i\dot v_i(t)-F_i(\vb*{x}(t),-\vb*{v}(t))^\dagger -\gamma_i v_i(t)]^2-\frac{1}{2m_i}\partial_{v_i}F_i(\vb*{x}(t),-\vb*{v}(t))^\dagger -\frac{\gamma_i}{2m_i}}).
\end{align}
Within the framework of stochastic thermodynamics \cite{Seifert.2012.RPP}, the total entropy production over the interval $[0,T]$ can be defined as the log-ratio of path probabilities:
\begin{equation}
	\Sigma \coloneqq \ev{ \ln \frac{\mds{P}(\Gamma)}{\widetilde{\mds{P}}(\widetilde{\Gamma})} }.
\end{equation}
This admits the following explicit expression in terms of the irreversible components of the probability currents:
\begin{equation}\label{eq:ent.prod.form}
	\Sigma=\int_0^T\dd{t}\iint\dd{\vb*{x}}\dd{\vb*{v}}\sum_{i=1}^d\frac{m_i^2}{\gamma_iT_i}\frac{J_{v_i}^{\rm irr}(\vb*{x},\vb*{v},t)^2}{p_t(\vb*{x},\vb*{v})}.
\end{equation}

We are interested in generalized currents $\phi(\Gamma)$, which satisfy the minimal time-antisymmetry condition, $\phi(\Gamma)=-\phi(\widetilde{\Gamma})$.
This class includes but is not limited to conventional time-integrated currents previously studied in the literature \cite{Fischer.2018.PRE,Vu.2019.PRE.UnderdampedTUR,Lee.2021.PRE.TUR}.
Our goal is to derive a universal bound on the relative fluctuation of such currents. As in the quantum case, we define an asymmetry term that captures the disparity between forward and backward dynamics:
\begin{equation}
	\Sigma_*\coloneqq\ev{\ln\frac{\widetilde{\mds{P}}(\widetilde{\Gamma})}{\mds{P}(\widetilde{\Gamma})}}=\ev{\ln\frac{\widetilde{\mds{P}}(\widetilde{\Gamma}|\vb*{x}(T),-\vb*{v}(T))p_T(\vb*{x}(T),\vb*{v}(T))}{\mds{P}(\widetilde{\Gamma}|\vb*{x}(T),-\vb*{v}(T))p_0(\vb*{x}(T),-\vb*{v}(T))}}.
\end{equation}
This term admits the following explicit representation:
\begin{align}
	\Sigma_*&=\int_0^T\dd{t}\sum_{i=1}^d\ev{\mca{G}_i(\vb*{x},\vb*{v},t)[F_i^-(\vb*{x},-\vb*{v})]}_t+\ev{\ln\frac{p_T(\vb*{x}(T),\vb*{v}(T))}{p_0(\vb*{x}(T),-\vb*{v}(T))}},\label{eq:asym.form}
\end{align}
where $\ev{\circ}_t\coloneqq\iint\dd{\vb*{x}}\dd{\vb*{v}}\circ p_t(\vb*{x},\vb*{v})$, $F_i^\pm(\vb*{x},\vb*{v})\coloneqq [F_i(\vb*{x},\vb*{v})^\dagger \pm F_i(\vb*{x},\vb*{v})]/2$, and the superoperators $\{\mca{G}_i\}$ are defined as
\begin{equation}
	\mca{G}_i(\vb*{x},\vb*{v},t)[\circ]\coloneqq \frac{1}{\gamma_iT_i}\qty{F_i(\vb*{x},\vb*{v})-F_i^+(\vb*{x},-\vb*{v})-2\gamma_iv_i-\frac{\gamma_iT_i}{m_i}\qty[\partial_{v_i}\ln p_t(\vb*{x},\vb*{v})] + \frac{\gamma_iT_i}{m_i}\partial_{v_i}}\circ.
\end{equation}
We further clarify how the term $\Sigma_*$ characterizes the forward-backward asymmetry. Notably, there are two primary sources of discrepancy between the forward and backward processes:
\begin{enumerate}
	\item Difference in force fields: The first source arises from the discrepancy between the force fields $\{F_i\}$ in the forward dynamics and $\{F_i^\dagger\}$ in the backward dynamics. In the presence of time-reversal-odd parameters (e.g., magnetic fields), these forces generally differ: $F_i \neq F_i^\dagger$. This asymmetry leads to fundamentally distinct forward and backward trajectories. When $F_i = F_i^\dagger$ for all $i$, the forward and backward dynamics become identical. This contribution is encapsulated in the function $F_i^- = (F_i^\dagger - F_i)/2$, which appears in the first term of Eq.~\eqref{eq:asym.form}.
	\item Difference in initial distributions: The second source stems from the difference in the initial probability distributions. For each time-reversed trajectory $\widetilde{\Gamma}$, the forward process assigns the initial probability $p_0(\vb*{x}(T), -\vb*{v}(T))$, while the backward process uses $p_T(\vb*{x}(T), \vb*{v}(T))$. This mismatch is captured in the second term of Eq.~\eqref{eq:asym.form}, which quantifies the statistical asymmetry between the start of the forward and backward evolutions. This term can also be also regarded as the analog of the boundary term $\mfr{b}$ in the quantum case.
\end{enumerate}
Combining the asymmetry $\Sigma_*$ with the entropy production $\Sigma$ and applying the same approach as in the quantum case, we arrive at the generalized TUR for underdamped Langevin dynamics:
\begin{equation}
	\frac{\mvar[\phi]}{\ev{\phi}^2}\ge f(\Sigma+\Sigma_*).
\end{equation}
This result reveals that both entropy production and forward-backward asymmetry jointly constrain the precision of generalized currents in general underdamped systems.

\subsection{Derivation of Eqs.~\eqref{eq:ent.prod.form} and \eqref{eq:asym.form}}
\sectionprl{Derivation of Eq.~\eqref{eq:ent.prod.form}}The total entropy production can be decomposed into the system and environmental components as
\begin{align}
	\Sigma&=\ev{\ln\frac{p_0(\vb*{x}(0),\vb*{v}(0))}{p_T(\vb*{x}(T),\vb*{v}(T))}}+\ev{\ln\frac{\mds{P}(\Gamma|\vb*{x}(0),\vb*{v}(0))}{\widetilde{\mds{P}}(\widetilde{\Gamma}|\vb*{x}(T),-\vb*{v}(T))}}\notag\\
	&\eqqcolon \Delta S+\Delta S_{\rm env}.
\end{align}
The system entropy change can be calculated as follows:
\begin{align}
	\Delta S&=\iint\dd{\vb*{x}}\dd{\vb*{v}}[p_0(\vb*{x},\vb*{v})\ln p_0(\vb*{x},\vb*{v}) -  p_T(\vb*{x},\vb*{v})\ln p_T(\vb*{x},\vb*{v})]\notag\\
	&=-\int_0^T\dd{t}\iint\dd{\vb*{x}}\dd{\vb*{v}}\dot p_t(\vb*{x},\vb*{v})\ln p_t(\vb*{x},\vb*{v})\notag\\
	&=\int_0^T\dd{t}\iint\dd{\vb*{x}}\dd{\vb*{v}}\sum_{i=1}^d\qty[\partial_{x_i}J_{x_i}(\vb*{x},\vb*{v},t)+\partial_{v_i}J_{v_i}(\vb*{x},\vb*{v},t)]\ln p_t(\vb*{x},\vb*{v})\notag\\
	&=-\int_0^T\dd{t}\iint\dd{\vb*{x}}\dd{\vb*{v}}\sum_{i=1}^d\qty[J_{x_i}(\vb*{x},\vb*{v},t)\partial_{x_i}\ln p_t(\vb*{x},\vb*{v}) + J_{v_i}(\vb*{x},\vb*{v},t)\partial_{v_i}\ln p_t(\vb*{x},\vb*{v})]\notag\\
	&=-\int_0^T\dd{t}\iint\dd{\vb*{x}}\dd{\vb*{v}}\sum_{i=1}^d\qty[ v_i\partial_{x_i}p_t(\vb*{x},\vb*{v}) + J_{v_i}^{\rm rev}(\vb*{x},\vb*{v},t)\partial_{v_i}\ln p_t(\vb*{x},\vb*{v})+J_{v_i}^{\rm irr}(\vb*{x},\vb*{v},t)\partial_{v_i}\ln p_t(\vb*{x},\vb*{v})]\notag\\
	&=-\int_0^T\dd{t}\iint\dd{\vb*{x}}\dd{\vb*{v}}\sum_{i=1}^d\qty[ \frac{1}{m_i} F_{i}^{\rm rev}(\vb*{x},\vb*{v})\partial_{v_i}p_t(\vb*{x},\vb*{v})+J_{v_i}^{\rm irr}(\vb*{x},\vb*{v},t)\partial_{v_i}\ln p_t(\vb*{x},\vb*{v})].
\end{align}
Similarly, using the fact $F_i(\vb*{x},-\vb*{v})^\dagger = F_i^{\rm rev}(\vb*{x},\vb*{v}) - F_i^{\rm irr}(\vb*{x},\vb*{v})$, the environmental entropy production can be calculated as
\begin{align}
	\Delta S_{\rm env}&=\ev{\ln\frac{\mds{P}(\Gamma|\vb*{x}(0),\vb*{v}(0))}{\widetilde{\mds{P}}(\widetilde{\Gamma}|\vb*{x}(T),-\vb*{v}(T))}}\notag\\
	&=\Bigg\langle\int_0^T\dd{t}\sum_{i=1}^d\Big[\frac{1}{4\gamma_iT_i}\Big\{[m_i\dot v_i(t)-F_i^{\rm rev}(\vb*{x}(t),\vb*{v}(t)) + F_i^{\rm irr}(\vb*{x}(t),\vb*{v}(t)) -\gamma_i v_i(t)]^2 \notag\\
	&\hspace{3cm}- [m_i\dot v_i(t)-F_i^{\rm rev}(\vb*{x}(t),\vb*{v}(t))-F_i^{\rm irr}(\vb*{x}(t),\vb*{v}(t))+\gamma_i v_i(t)]^2\Big\} - \frac{1}{m_i}\partial_{v_i}F_i^{\rm rev}(\vb*{x}(t),\vb*{v}(t))\Big]\Bigg\rangle\notag\\
	&=\ev{\int_0^T\dd{t}\sum_{i=1}^d\qty{\frac{1}{\gamma_iT_i}\qty[F_i^{\rm irr}(\vb*{x}(t),\vb*{v}(t)) -\gamma_i v_i(t)]\qty[m_i\dot v_i(t)-F_i^{\rm rev}(\vb*{x}(t),\vb*{v}(t))] - \frac{1}{m_i}\partial_{v_i}F_i^{\rm rev}(\vb*{x}(t),\vb*{v}(t))}}\notag\\
	&=\int_0^T\dd{t}\iint\dd{\vb*{x}}\dd{\vb*{v}}\sum_{i=1}^d\qty{\frac{1}{\gamma_iT_i}[F_i^{\rm irr}(\vb*{x},\vb*{v})-\gamma_i v_i][m_iJ_{v_i}(\vb*{x},\vb*{v},t)-F_i^{\rm rev}(\vb*{x},\vb*{v})p_t(\vb*{x},\vb*{v})] - \frac{1}{m_i}p_t(\vb*{x},\vb*{v})\partial_{v_i}F_i^{\rm rev}(\vb*{x},\vb*{v})}\notag\\
	&=\int_0^T\dd{t}\iint\dd{\vb*{x}}\dd{\vb*{v}}\sum_{i=1}^d\qty{\frac{m_i}{\gamma_iT_i}J_{v_i}^{\rm irr}(\vb*{x},\vb*{v},t)\qty[F_i^{\rm irr}(\vb*{x},\vb*{v})-\gamma_i v_i]+\frac{1}{m_i}F_i^{\rm rev}(\vb*{x},\vb*{v})\partial_{v_i}p_t(\vb*{x},\vb*{v})}.
\end{align}
Consequently, the total entropy production can be expressed as
\begin{align}
	\Sigma&=\Delta S+\Delta S_{\rm env}\notag\\
	&=\int_0^T\dd{t}\iint\dd{\vb*{x}}\dd{\vb*{v}}\sum_{i=1}^d J_{v_i}^{\rm irr}(\vb*{x},\vb*{v},t)\qty{\frac{m_i}{\gamma_iT_i}\qty[F_i^{\rm irr}(\vb*{x},\vb*{v})-\gamma_i v_i] - \partial_{v_i}\ln p_t(\vb*{x},\vb*{v})}\notag\\
	&=\int_0^T\dd{t}\iint\dd{\vb*{x}}\dd{\vb*{v}}\sum_{i=1}^d\frac{m_i^2}{\gamma_iT_i}\frac{J_{v_i}^{\rm irr}(\vb*{x},\vb*{v},t)^2}{p_t(\vb*{x},\vb*{v})},
\end{align}
which is exactly Eq.~\eqref{eq:ent.prod.form}.

\sectionprl{Derivation of Eq.~\eqref{eq:asym.form}}The asymmetry can be decomposed into two contributions as
\begin{equation}
	\Sigma_*=\ev{\ln\frac{\widetilde{\mds{P}}(\widetilde{\Gamma}|\vb*{x}(T),-\vb*{v}(T))}{\mds{P}(\widetilde{\Gamma}|\vb*{x}(T),-\vb*{v}(T))}}+\ev{\ln\frac{p_T(\vb*{x}(T),\vb*{v}(T))}{p_0(\vb*{x}(T),-\vb*{v}(T))}}.
\end{equation}
The first contribution can be calculated as follows:
\begin{align}
	&\ev{\ln\frac{\widetilde{\mds{P}}(\widetilde{\Gamma}|\vb*{x}(T),-\vb*{v}(T))}{\mds{P}(\widetilde{\Gamma}|\vb*{x}(T),-\vb*{v}(T))}}\notag\\
	&=\ev{-\int_0^T\dd{t}\sum_{i=1}^d\qty{\frac{1}{4\gamma_iT_i}[m_i\dot v_i(t)-F_i(\vb*{x}(t),-\vb*{v}(t))^\dagger -\gamma_i v_i(t)]^2-\frac{1}{2m_i}\partial_{v_i}F_i(\vb*{x}(t),-\vb*{v}(t))^\dagger -\frac{\gamma}{2m_i}}} \notag\\
	&+\ev{\int_0^T\dd{t}\sum_{i=1}^d\qty{\frac{1}{4\gamma_iT_i}[m_i\dot v_i(t)-F_i(\vb*{x}(t),-\vb*{v}(t))-\gamma_i v_i(t)]^2-\frac{1}{2m_i}\partial_{v_i}F_i(\vb*{x}(t),-\vb*{v}(t))-\frac{\gamma_i}{2m_i}}}\notag\\
	&=\ev{\int_0^T\dd{t}\sum_{i=1}^d\qty{\frac{1}{\gamma_iT_i} F_i^-(\vb*{x}(t),-\vb*{v}(t))[ m_i\dot v_i(t) - F_i^+(\vb*{x}(t),-\vb*{v}(t)) -\gamma_iv_i(t) ] + \frac{1}{m_i}\partial_{v_i}F_i^-(\vb*{x}(t),-\vb*{v}(t)) }}\notag\\
	&=\int_0^T\dd{t}\iint\dd{\vb*{x}}\dd{\vb*{v}}\sum_{i=1}^d\qty{\frac{m_i}{\gamma_iT_i}J_{v_i}(\vb*{x},\vb*{v},t)-\frac{1}{\gamma_iT_i}\qty[F_i^+(\vb*{x},-\vb*{v})+\gamma_iv_i]p_t(\vb*{x},\vb*{v}) + \frac{1}{m_i}p_t(\vb*{x},\vb*{v})\partial_{v_i}}F_i^-(\vb*{x},-\vb*{v})\notag\\
	&=\int_0^T\dd{t}\sum_{i=1}^d\ev{\mca{G}_i(\vb*{x},\vb*{v},t)[F_i^-(\vb*{x},-\vb*{v})]}_t.
\end{align}

\section{Propositions}
\begin{proposition}\label{prop:f.equiv}
The following equality holds for arbitrary positive $x$:
\begin{equation}
	f(x)=\csch^2[\Phi(x/2)].
\end{equation}
\begin{proof}
Since $f(x)=4[\Phi(x/2)/x]^2-1$ and $\csch x=2/(e^x-e^{-x})$, we need only prove that
\begin{equation}
	4\qty[\frac{\Phi(x/2)}{x}]^2-1=\frac{4}{[e^{\Phi(x/2)}-e^{-\Phi(x/2)}]^2}.
\end{equation}
This is equivalent to showing
\begin{align}
	\frac{\Phi(x/2)}{x/2}=\frac{e^{\Phi(x/2)}+e^{-\Phi(x/2)}}{e^{\Phi(x/2)}-e^{-\Phi(x/2)}}.
\end{align}
Since $x/2=\Phi(x/2)\tanh[\Phi(x/2)]$, this equality can be verified as follows:
\begin{align}
	\frac{\Phi(x/2)}{x/2}&=\frac{1}{\tanh[\Phi(x/2)]}\notag\\
	&=\frac{e^{\Phi(x/2)}+e^{-\Phi(x/2)}}{e^{\Phi(x/2)}-e^{-\Phi(x/2)}}.
\end{align}
This completes the proof.
\end{proof}
\end{proposition}